\documentclass[sigconf, nonacm]{acmart}

\newcommand\vldbdoi{XX.XX/XXX.XX}
\newcommand\vldbpages{XXX-XXX}
\newcommand\vldbvolume{14}
\newcommand\vldbissue{1}
\newcommand\vldbyear{2020}
\newcommand\vldbauthors{\authors}
\newcommand\vldbtitle{\shorttitle} 
\newcommand\vldbavailabilityurl{URL_TO_YOUR_ARTIFACTS}
\newcommand\vldbpagestyle{plain} 

\newtheorem{definition}{Definition}
\newtheorem{theorem}{Theorem}
\newtheorem{corollary}{Corollary}
\usepackage{subfigure}
\usepackage{enumitem}
\usepackage{algorithm}  
\usepackage[noend]{algorithmic} 
\usepackage{pifont}
\usepackage{multirow}
\usepackage{textcomp}
\usepackage{diagbox}
\usepackage{tabularray}
\usepackage{needspace}

\newcommand{\subject}[1]{{\textbf{#1}}}

\begin{document}
\title{$\boldsymbol{Steiner}$-Hardness: A Query Hardness Measure for \\Graph-Based ANN Indexes}

\author{Zeyu Wang}
\orcid{0000-0002-0455-0830}
\affiliation{%
  \institution{Fudan University}
}
\email{zeyuwang23@m.fudan.edu.cn}

\author{Qitong Wang}
\orcid{0000-0001-6360-3800}
\affiliation{%
  \institution{LIPADE, Universit{\'e} Paris Cit{\'e}}
}
\email{qitong.wang@etu.u-paris.fr}

\author{Xiaoxing Cheng}
\affiliation{%
  \institution{Tongji University}
}
\email{vozeo@tongji.edu.cn}

\author{Peng Wang}
\orcid{0000-0002-8136-9621}
\affiliation{%
  \institution{Fudan University}
}
\email{pengwang5@fudan.edu.cn}

\author{Themis Palpanas}
\orcid{0000-0002-8031-0265}
\affiliation{%
  \institution{LIPADE, Universit{\'e} Paris Cit{\'e}}
}
\email{themis@mi.parisdescartes.fr}

\author{Wei Wang}
\orcid{0000-0003-0264-788X}
\affiliation{%
  \institution{Fudan University}
}
\email{weiwang1@fudan.edu.cn}

\begin{abstract}
Graph-based indexes have been widely employed to accelerate approximate similarity search of high-dimensional vectors.
However, the performance of graph indexes to answer different queries varies vastly, leading to an unstable quality of service for downstream applications.
This necessitates an effective measure to test query hardness on graph indexes.
Nonetheless, popular distance-based hardness measures like LID lose their effects due to the ignorance of the graph structure.
In this paper, we propose $Steiner$-hardness, a novel connection-based graph-native query hardness measure.
Specifically, we first propose a theoretical framework to analyze the minimum query effort on graph indexes and then define $Steiner$-hardness as the minimum effort on a representative graph.
Moreover, we prove that our $Steiner$-hardness is highly relevant to the classical Directed $Steiner$ Tree (DST) problems.
In this case, we design a novel algorithm to reduce our problem to DST problems and then leverage their solvers to help calculate $Steiner$-hardness efficiently.
Compared with LID and other similar measures, $Steiner$-hardness shows a significantly better correlation with the actual query effort on various datasets.
Additionally, an unbiased evaluation designed based on $Steiner$-hardness reveals new ranking results, indicating a meaningful direction for enhancing the robustness of graph indexes. This paper is accepted by PVLDB 2025.
\end{abstract}

\maketitle

\pagestyle{\vldbpagestyle}
\begingroup\small\noindent\raggedright\textbf{PVLDB Reference Format:}\\
\vldbauthors. \vldbtitle. PVLDB, \vldbvolume(\vldbissue): \vldbpages, \vldbyear.\\
\href{https://doi.org/\vldbdoi}{doi:\vldbdoi}
\endgroup
\begingroup
\renewcommand\thefootnote{}\footnote{\noindent
This work is licensed under the Creative Commons BY-NC-ND 4.0 International License. Visit \url{https://creativecommons.org/licenses/by-nc-nd/4.0/} to view a copy of this license. For any use beyond those covered by this license, obtain permission by emailing \href{mailto:info@vldb.org}{info@vldb.org}. Copyright is held by the owner/author(s). Publication rights licensed to the VLDB Endowment. \\
\raggedright Proceedings of the VLDB Endowment, Vol. \vldbvolume, No. \vldbissue\ %
ISSN 2150-8097. \\
\href{https://doi.org/\vldbdoi}{doi:\vldbdoi} \\
}\addtocounter{footnote}{-1}\endgroup

\ifdefempty{\vldbavailabilityurl}{}{
\vspace{.3cm}
\begingroup\small\noindent\raggedright\textbf{PVLDB Artifact Availability:}\\
The source code, data, and/or other artifacts have been made available at \url{https://github.com/DSM-fudan/Steiner-hardness}.
\endgroup
}

\section{Introduction}
Approximate Nearest Neighbor (ANN) search has recently gained high importance. 
Compared to exact search, ANN can provide high-quality approximate answers at a significantly reduced query time.
To achieve this, ANN indexes are used, which are built before querying.
Among the various ANN index families, graph indexes\footnote{In this paper, we use \emph{graph indexes} rather than graph-based indexes for simplicity.} like HNSW~\cite{hnsw} have become the state-of-the-art for in-memory ng-approximate (i.e., with no guarantees for the accuracy of the results~\cite{DBLP:journals/pvldb/EchihabiZPB18}) search~\cite{hydra2,tkde-survey,ann-benchmark}.
In graph indexes, vectors are represented as vertices in a graph, and connected by edges based on some kind of proximity between vectors.
During query answering, the algorithm starts from an entry vertex and travels along the edges to compute the $k$NN answer.

\begin{figure}[tb]
\includegraphics[width=\linewidth]{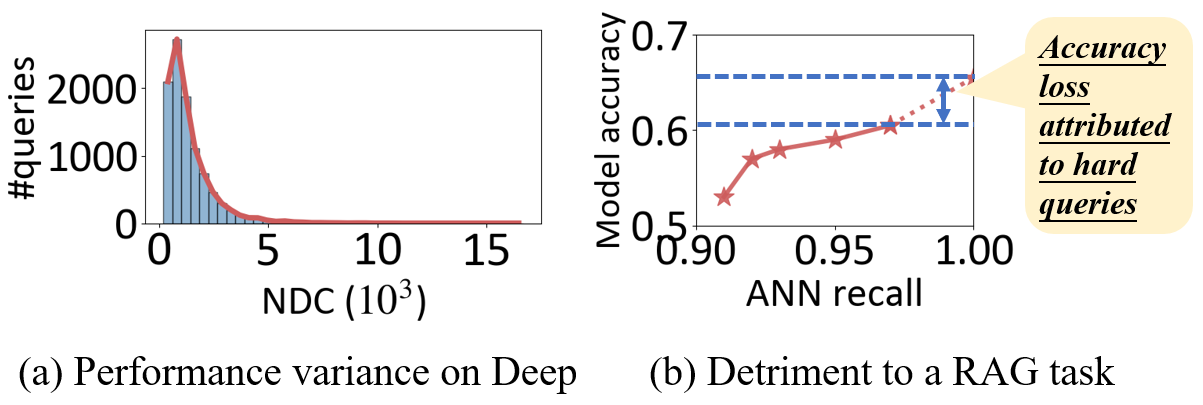}
\caption{Query performance variance on graph indexes. (a) Histograms of NDC to reach 90\% recall on Deep~\cite{deep} dataset. (b) A real example on a RAG task~\cite{rag}, where the low recall of hard queries impairs model accuracy.}
\label{fig:motivation} 
\end{figure}

Although graph indexes show superior query performance, it is observed that the performance of different queries varies vastly~\cite{early-termination,martin-survey,lid}.
That is, graph indexes answer some queries efficiently as expected (i.e., for simple queries), while for other queries, they perform much worse (i.e., for hard queries).
This problem widely occurs in various datasets.
As shown in Figure~\ref{fig:motivation}(a), to reach 90\% recall, the Number of Distance Calculations (NDC)\footnote{NDC is widely used in the literature~\cite{ms-benchmark,spann,bulletin-wang} as a machine-independent indicator of query time.} of the queries varies by 3 orders of magnitude 
for the Deep dataset~\cite{deep}.

Such a large variance in performance will impair the downstream tasks.
For example, in recommendation systems, a reduced search accuracy might drift the user away since irrelevant products are recommended to them, if their profiles correspond to hard queries~\cite{rec1,rec2,rec3}.
Another example is shown in Figure~\ref{fig:motivation}(b), where graph indexes are used in a Retrieval-Augmented Generation (RAG) framework~\cite{rag}.
Specifically, HNSW index retrieves related items from documents, and these items act as the prompt for language models.
In this case, the recall of ANN search influences model accuracy.
In a careful parameter setting, the average recall is 0.97; yet, some hard queries suffer from recall below 0.2, leading to suboptimal answers by the language model.
Note that in actual applications the frequency of such hard queries is not known, and a reliable ANN index is required to be able to effectively handle all queries.
In this context, an essential problem is required to be addressed: \emph{How can we design a hardness measure for graph indexes to differentiate simple from hard queries?}

\subject{The problem of LID.}
In the past years, Local Intrinsic Dimensionality (LID)~\cite{lid} has become the most widely used measure to gauge the difficulty of queries and datasets.
Formally, given a query $q$, LID at distance $r$ is defined as  
$
LID_q(r) = \lim_{\epsilon\rightarrow0^+} \\ \frac{ln(F((1+\epsilon)r)/ln(F(r)))}{ln(1+\epsilon)},
$
where $F$: $\mathbb{R}\rightarrow[0,1]$ is the cumulative distribution function of distances to $q$.
When using LID as the hardness, $r$ is set to be the distance between $q$ and the $k$-th NN.
LID describes the increasing rate of mass w.r.t. the radius at some radius $r$.
The more densely data are distributed at $k$NN neighborhood, the larger $LID$ is which indicates a harder query.

Generally speaking, LID, along with other existing hardness measures~\cite{epsilon,qe,rc}, describes the hardness in terms of the data distribution in the high-dimensional space.
It works well on partition-based indexes~\cite{epsilon,epsilon2,lid} since it evaluates how hard it is to distinguish $k$NN and other points w.r.t. the distance to the query.
However, on graph indexes, the effort to answer a query is directly determined by the connections (i.e., edges), rather than the distance.
Thus, for the points that are close to the query but not in the $k$NN answer, if they are not located on the path from the entry point towards the $k$NN answers, they will never be accessed during querying.

\begin{figure}[tb]
  \includegraphics[width=\linewidth]{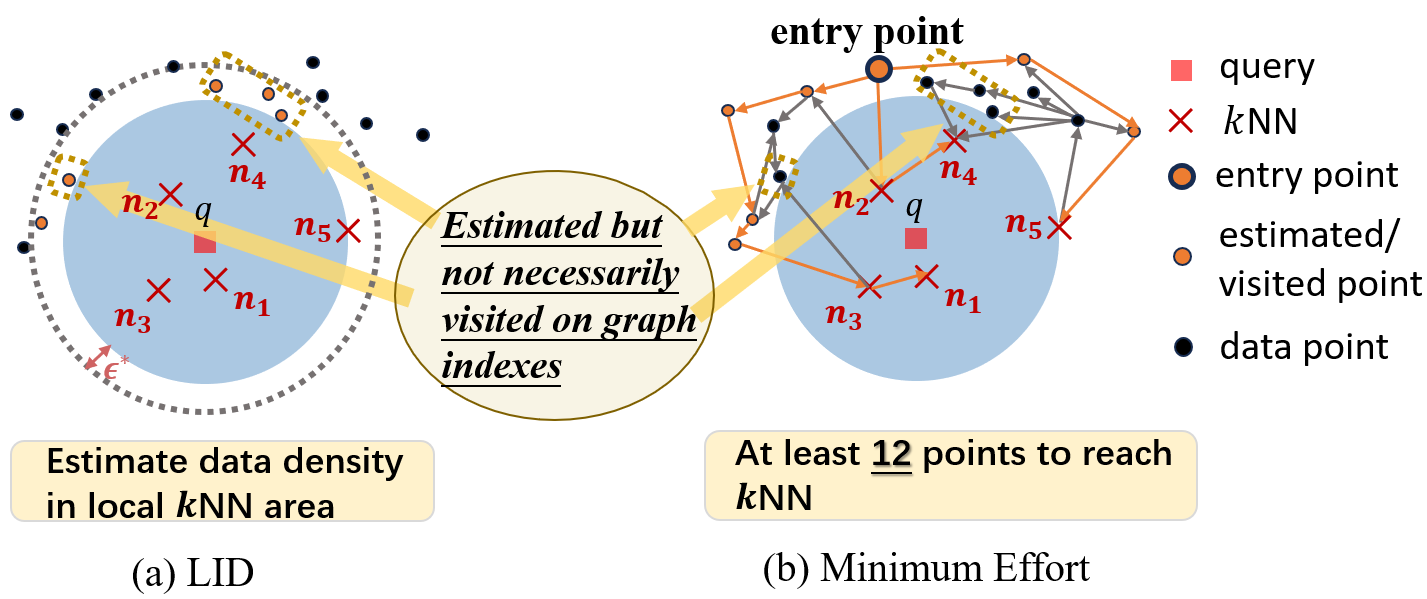}
  \caption{Comparison of LID and ME on the same dataset.}
  \label{fig:llus}
\end{figure}

Figure~\ref{fig:llus} shows an illustrative example of measuring the hardness of a query $q$ with $k$=5, where $n_1$ to $n_5$ are the five NNs.
We show the principle of LID in Figure~\ref{fig:llus}(a), where the number of points with slightly larger distance than $n_5$ to $q$ (i.e., orange points) are also considered for hardness estimation.
However, Figure~\ref{fig:llus}(b) shows the shortest path from an entry point to all the NNs.
Using the given graph connections, the shortest path will not access the orange points of Figure~\ref{fig:llus}(a).
In other words, the point density in the $k$NN neighborhood does not necessarily influence the hardness of a query in a graph index.
This raises the need for a graph-native connection-based hardness measure.
Such a measure should be designed based on the query effort analysis on the graph (rather than on the data distribution, like LID).

To achieve this goal, there are three major research challenges that need to be tackled in order:
\begin{enumerate}[noitemsep, topsep=0pt, wide=\parindent]
    \item \emph{(Query-Algorithm-Independent Effort Analysis) Given graph index $G$ and query $q$, how can we estimate the (minimum) effort to answer $q$ in $G$?}
    \item \emph{(Index-Independent Query Hardness Measure) How can we judge how hard it is to answer a query $q$ with some graph index?} 
    \item \emph{(Efficiency) How can we calculate the hardness efficiently?}
\end{enumerate}

\subject{Query Effort Analysis.}
To analyze the query effort on graph indexes, many theoretical studies have been proposed recently~\cite{nsg,nssg,tau,theory-understand,theory,theory2}.
However, there are still four major limitations that render them impractical.
First, most studies~\cite{nsg,nssg,tau,theory,theory2} assume a uniform data distribution, which is very far from the distributions observed in real datasets.
Second, the graph structures considered in these studies, such as MSNET~\cite{nsg,nssg,tau}, $\rho$-Graph~\cite{theory,theory2}, and exact Vamana~\cite{worst-case}, differ from the graph indexes used in practice. 
Third, the expected- or worst-case results are insufficient to explain the behaviors observed in practice~\cite{note}.
Last but not least, all these studies focus on an over-simplified single-direction (i.e., $ef$=1) greedy search algorithm for the \textbf{1}NN problem.

\begin{figure}[tb]
\subfigure[LID]{
  \includegraphics[width=0.475\linewidth]{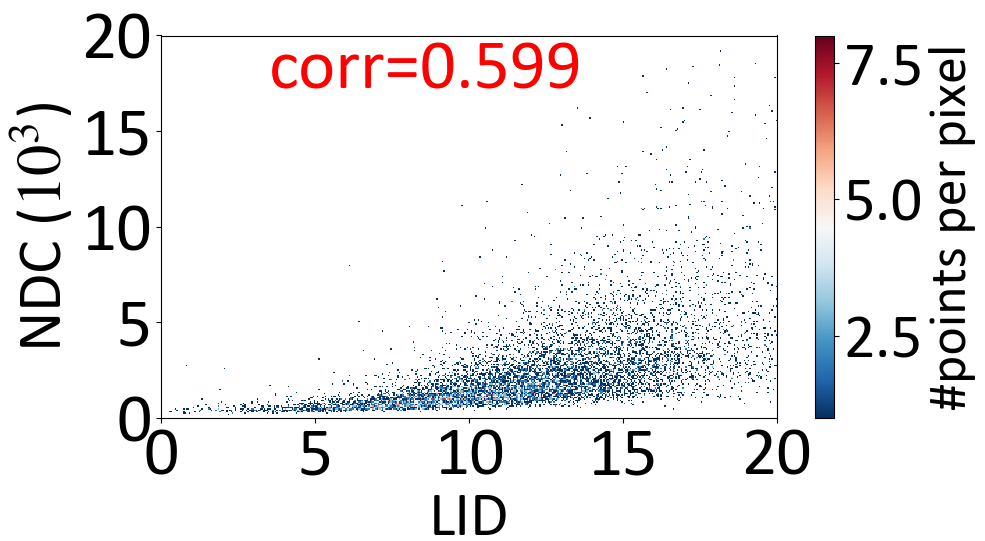}
}
\subfigure[$Steiner$-hardness]{
\includegraphics[width=0.475\linewidth]{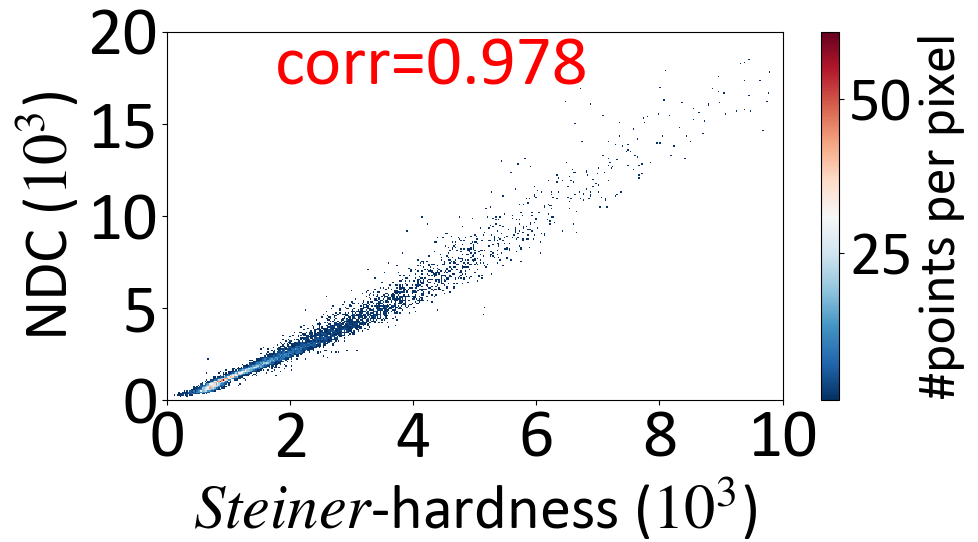}
}
\caption{The correlation between $Steiner$-hardness (b) and NDC to reach 90\% recall is much stronger than LID (a).}
\label{fig:correlation} 
\end{figure}

To overcome these limitations, we propose a theoretical framework to analyze the \emph{Minimum Effort} (ME) to answer queries using a graph index.
Our framework introduces a lower bound of the real effort to answer a query using a given graph, based only on the connections of that graph.
That is, ME represents the effort of an optimal query algorithm on the given graph. 
Our framework incorporates three major improvements.
First, 
our analysis can be applied to any graph index and any data distribution.
Second, our framework is designed to describe the query effort under different \emph{recall targets}, which is important for the ANN problem.
Third, our framework focuses on the analysis of the real bottleneck in the query answering process: as we explain in Section~\ref{sec:me}, most of the query time is spent on identifying the rest of the $k$NN answers after having identified the first of the $k$NN answers. 
However, this part of the query answering process has not been studied in detail in the past.
With these improvements, our theoretical framework can effectively estimate the query effort on a given graph index and explain the performance variance.

\subject{Query Hardness Measure.}
To further define an index-independent query hardness measure based on the ME, it is important to extract the common feature of the current indexes.
We observe that current graph indexes, even though are of various types, have two common structural features.
First, most of the edges are short-range connections.
They can all be roughly viewed as subgraphs of a large $K$Graph.
Second, they use edge-pruning rules to control the out-degree and sparsify the graph, in expectation of a better navigability-sparsity tradeoff~\cite{wang-survey}.
To this effect, we select MRNG (Monotonic Relative Neighborhood Graph)~\cite{nsg} as the representative graph, which prunes edges from short links with direction-based pruning rules.
Then we define our hardness measure, $Steiner$-hardness, as the ME on this MRNG.
Figure~\ref{fig:correlation} displays the correlation between LID, $Steiner$-hardness and the NDC to reach 90\% recall for the queries on Deep dataset with HNSW.
We observe that the correlation is significantly improved with our $Steiner$-hardness, which demonstrates the effectiveness of the proposed measure.
\subject{Efficient algorithms to calculate $\boldsymbol{Steiner}$-hardness.}
Although $Steiner$-hardness is effective for measuring the query difficulty, it can be computationally intensive when implemented in a naive manner.
To overcome this problem, we first draw the connections between ME and the classical Directed $Steiner$ Tree (DST) problem~\cite{dst} and its variants~\cite{compendium}, and then leverage the DST solvers to solve ME efficiently. 
Moreover, we design a novel union-find set based algorithm to reduce our ME to the standard DST problems, which provides more than 1000x speedup when compared to the naive method.

\subject{Unbiased Workloads Generation.}
Finally, we observe that the hardness distribution of current query workloads is biased. 
As shown in Figure~\ref{fig:motivation}, most of the queries are simple, and only a small minority are hard.
When using the average performance to show the ability of an index, the result will be dominated by the simple queries, and thus, become over-optimistic.
Unfortunately, current benchmarks do not pay particular attention to the distribution of the query workloads~\cite{wang-survey,tkde-survey,ann-benchmark,ms-benchmark}.
To tackle this problem, we propose a method to build an unbiased workload where queries fall within the same distribution as the dataset, and the hardness of queries follows a uniform distribution.
Using this workload, we can stress-test the graph indexes, and evaluate their performance when faced with queries of various hardness.

Our contributions can be summarized as follows. 

(1) We propose a theoretical framework to analyze the practical effort to answer approximate $k$NN queries on graph indexes,
which effectively estimates the actual query effort (Section~\ref{sec:me}).

(2) We develop a novel $Steiner$-hardness measure for queries on graph indexes based on our theoretical framework, along with an efficient algorithm to calculate it with the help of DST solvers.
$Steiner$-hardness effectively differentiates the difficulty of different queries for graph indexes (Sections~\ref{sec:alg-delta} and ~\ref{sec:hardness}).

(3) We propose a method for building unbiased workloads with variable hardness. 
This method can then be used to evaluate the comprehensive performance of graph indexes (Section~\ref{sec:workload-gen}).

(4) We conduct extensive experiments on public datasets to verify the effectiveness of the proposed analytical framework, $Steiner$-hardness and workload generation methods. 
Moreover, we evaluate current graph indexes with the new unbiased query workloads.
The result provides new insights for index selection (Section~\ref{sec:expr}).

\section{Preliminaries}
\label{sec:pre}

\subsection{Problem Setting}

In this paper, we assume a high-d vector database $db$ containing $N$ vectors of length $d$.
The exact $k$NN query can be defined as follows.
\begin{definition}[$k$NN Query]
Given an integer $k$, a query vector $q$ of dimensionality $d$, and a distance measure $D$, a \textbf{$k$NN query} retrieves from $db$ the set of vectors $N_k=\{n_1,n_2,\dots,n_k\}$ such that for any other vector $v$ in $db$ and any $n_i \in N_k$, $D(n_i,q) \leq D(v,q)$. 
\end{definition}

For simplicity, we denote the distance between query and the $i$-th nearest neighbor by $d_i=D(q,n_i)$.
Approximate $k$NN query returns approximate answers, $AN_k=\{an_1,an_2,\dots,an_k\}$, instead of the exact $k$NN results.
Recall is often used to indicate the quality of the approximate answers.
Formally, 
$
    Recall@k = \frac{|AN_k \cap N_k|}{k}.
$

\subsection{Query on Graph Indexes}
\label{sec:graph-intro}

\begin{algorithm}[tb]
\caption{Greedy search (graph $G$, query $q$, entry point $ep$, $ef$)} 
\label{alg:query}
{\scriptsize
\begin{algorithmic}[1]
\STATE $pq$ = a priority queue with unlimited capacity, initialized with $ep$
\STATE $H$  = a max-heap with capacity $ef$ 
\WHILE{$pq$ is not empty}
    \STATE $d_{v_c}, v_c$ = pop an element from $pq$
    \STATE $d_{v_{top}}, v_{top}$ = the heap top of $H$
    \IF{$d_{v_c} > d_{v_{top}}$}
        \STATE break
    \ENDIF
    \FOR{each neighbor $v$ of $v_c$ which has not been accessed}
        \IF{$D(v, q) < d_{v_{top}}$}
            \STATE Insert $(D(v, q), v)$ into $pq$ and $H$
        \ENDIF
        \STATE mark $v$ as accessed
    \ENDFOR
    \STATE resize $H$ to be $ef$
\ENDWHILE
\RETURN $k$ smallest elements in $H$
\end{algorithmic}
} 
\end{algorithm}

Graph indexes use a directed graph $G(V,E)$ to index vectors, where each vector is represented as a vertex in $V$, and the edges in $E$ connect vectors based on some kind of proximity.
Graph indexes commonly use greedy search to retrieve $k$NN.
As shown in Algorithm~\ref{alg:query} and Figure~\ref{fig:llus}(c), the search starts from an entry point $ep$, which is often selected randomly, and then computes the distance between the neighbors of $ep$ to the query $q$.
The accessed points are stored in a priority queue $pq$.
In the next step, the algorithm selects the closest point to $q$ from $pq$ as the next stop to visit and repeats the above process.
Note that not all accessed points can enter $pq$: the algorithm maintains a size-bounded heap $H$ and only the points that are closer than some point in $H$ are qualified to enter $pq$.
Finally, the algorithm terminates when all the points in $pq$ are farther than the points in $H$ to $q$.
The algorithm is greedy because only relatively close points to the query can be accessed.
For example, a point of $N_k$ who is the neighbor of a distant point to $q$ will not be accessed.
A direct way to escape such ``local optimum'' is to increase the capacity of $H$, i.e., $ef$, which is also the knob to tune the efficiency-accuracy trade-off in graph indexes.

\section{Related Work}
\label{sec:related-work}
\subsection{ANN indexes}
\subject{Partition-based indexes.} Partition-based indexes usually first reduce the dimensions of vectors 
and then build indexes to partition the lower-dimensional space.
When querying, the index only accesses data in the partitions that are near to the query.
These indexes often provide a quality guarantee for approximate queries~\cite{hydra2}.
Several index families are designed based on this rationale.
The Locality Sensitive Hashing (LSH) indexes~\cite{pmlsh,lsh,qalsh,c2lsh,r2lsh} use hash functions to project high-d vectors and build spatial indexes.
Some tree indexes~\cite{DBLP:journals/vldb/PengFP21,DBLP:conf/icde/PengFP21,DBLP:journals/pvldb/ChatzakisFKPP23,dumpy,hercules,ds-tree,dumpyos,civet} use summarization techniques such as iSAX~\cite{isax} to reduce dimensionality and build ad-hoc tree indexes to partition the space.
Others directly partition the space hierarchically with hyper-planes~\cite{annoy,rpt-tree,lanns}.
Pivot-based indexes~\cite{pivot,hd-index,flann,spann,spb-tree,pq,rabitq} cluster data according to the distance to a given group of pivots~\cite{pivot-select}.

\subject{Graph indexes.} In the past decade, the superior query performance of graph indexes has attracted great interest from the research and industrial~\cite{milvus,manu,adbv,spfresh,vbase} communities.
The simplest graph index is $K$Graph~\cite{kgraph}, where each vector connects to its $K$ nearest neighbors.
Inspired by the small-world phenomenon~\cite{small-world}, NSW~\cite{nsw} adds more long-range links to achieve better navigability, and HNSW~\cite{hnsw} uses the RNG rule to sparsify the graph and bounds the out-degree of each point.
In this way, HNSW achieves a balance between sparsity and navigability leading to a significant improvement.
DPG~\cite{tkde-survey}, NSG~\cite{nsg} and NSSG~\cite{nssg} further study the influence of the angle between edges on search performance when sparsifying the graph.
$\tau$-MNG~\cite{tau}, DEG~\cite{k-regular} and ~\cite{fanng,hcnng,grasp} also study the navigability-sparsity tradeoff in theory and practice.
Some works focus on the selection of entry points~\cite{lsh-apg,hvs} while others optimize the search process~\cite{early-termination,stronger,leqat,adsampling,finger,dco}.
The evolution of the graph index design is discussed and verified in~\cite{bulletin-wang}.
Wang et al.~\cite{wang-survey} conducts a comprehensive survey on graph indexes.

\subsection{Hardness Measures}
A few hardness measures are leveraged to measure the query hardness.
The most popular measure is \textbf{LID}~\cite{lid,lid-def,lid-def2}.
Besides LID, there are several other local intrinsic dimension models~\cite{ed,ged,mnd} following a similar rationale.
In this paper, we select the most popular LID as in ~\cite{lid}.
Due to the computational difficulty, LID is usually estimated by the following equation, which is a maximum-likelihood estimation~\cite{id,mle}. Formally, $\hat{LID_q} = -(\frac{1}{k}\sum_{i=1}^kln\frac{d_i}{d_k})^{-1}$.
Relative Contrast (\textbf{RC})~\cite{rc} and Query Expansion (\textbf{QE})~\cite{qe} are also studied in ~\cite{lid} as hardness measures based on data distribution.
RC presents a global view for other dataset points by using $d_{mean}$ as an indicator while QE considers the density in the local area.
Formally, $RC_q=\frac{d_{mean}}{d_k}$ where $d_{mean}$ is the mean distance between $q$ and all the vectors in $db$, and $QE_q=\frac{d_{2k}}{d_k}$.
Larger RC and QE mean simpler queries.
$\boldsymbol{\epsilon}$\textbf{-hardness}
is also proposed for pruning-based tree indexes~\cite{epsilon,epsilon2}.
Given the pruning ability of an index on a specific query, there are some close points to the query that cannot be pruned.
These points are viewed as the ME to answer this query. Formally, $\epsilon$-hardness of a query $q$ is defined as $|\{v|v \in db \wedge D(q,v) \leq (1+\epsilon)d_k\}|$, where $\epsilon$ is a user-determined parameter.
A recent study predicts the hardness on-the-fly, while the query is executed halfway, using learned models based on the found best-so-far answers~\cite{early-termination}.
However, it cannot reveal the intrinsic hardness.

\subsection{Theoretical Study of Query Complexity on Graph Indexes}
\label{sec:relate-theory}
Although graph indexes are proposed based on heuristics~\cite{small-world}, there are also some studies trying to analyze the complexity of query answering on graph indexes.
The concept of MSNET~\cite{msnet} is introduced in ~\cite{nsg} and the expected length of the search path between two points in the dataset in MSNET is $\mathcal{O}(n^{\frac{2}{d}}logn)$ when using a simplified greedy search algorithm under the uniform distribution.
Moreover, it proposes a prototype MRNG which is an instance of MSNET.
MRNG modifies the RNG~\cite{rng} to be a directed graph.
On MRNG, a point will link to its nearest neighbors unless the neighbors are located in a very close direction.
(More details provided in Section~\ref{sec:hardness}.)
The case when the query is not in the dataset is discussed in ~\cite{nssg}.
$\tau$-MG~\cite{tau} is proposed to optimize MRNG when $d_1<\tau$ while in ~\cite{theory-understand} MRNG is generalized by bounding the out-degree and the considered candidates in the building stage.
~\cite{theory,theory2} study this problem on $\rho$-Graph of different ranges of dimensionality.
The worst-case analyses are first studies in ~\cite{worst-case} under datasets with bounded doubling dimension on exact Vamana graph~\cite{diskann}.
However, this study is still limited in $k$=1 and the simplified greedy search algorithm
and the handcrafted hard queries somewhat deviate from the original data distribution.
Delaunay graph~\cite{dg} is sometimes utilized to explain this~\cite{wang-survey,nsg,tau},
which however, quickly becomes a complete graph as $d$ grows.
Recently, ~\cite{note} tries to build connections between the size of strongly connected components of $k$NN-induced subgraph and the recall of a query, which inspires our work.

\begin{figure*}[tb]
  \includegraphics[width=\linewidth]{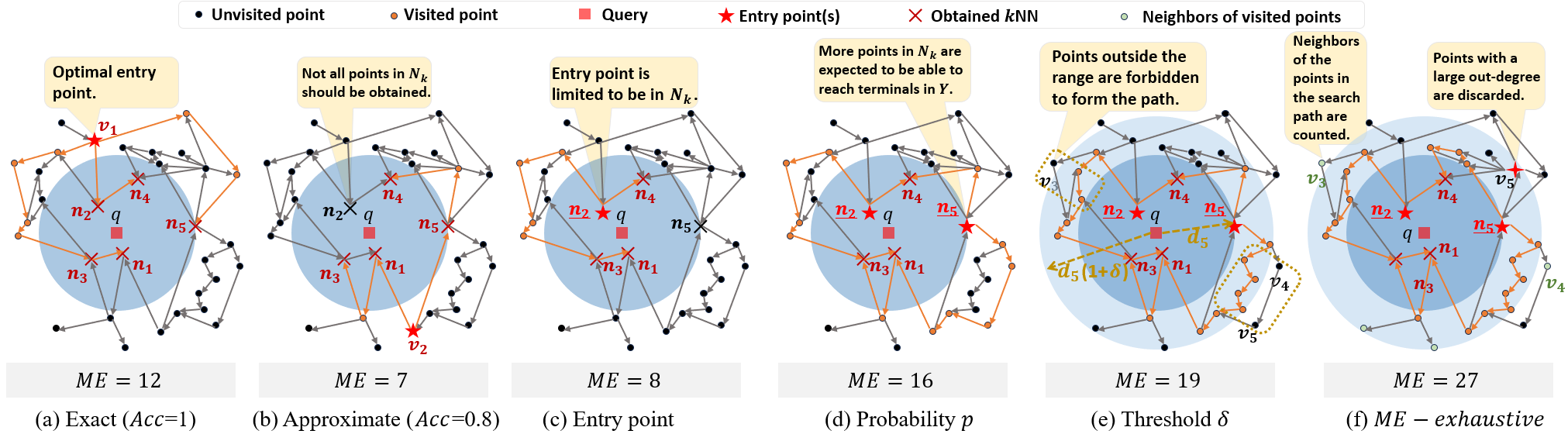}
  \caption{An illustrative example of our ME definitions with $k$=5. The orange points and edges form $Y$. (a) $Acc$=100\%, (b) $Acc$=80\%, (c) entry point is limited to be in $N_k$, (d) $p$=0.4, (e) Limited range of candidates, (f) $ME-exhaustive$. }
  \label{fig:me-illus}
\end{figure*}


\section{Characterize the Effort of Answering a Query}
\label{sec:me}
In this section, we analyze the effort to answer approximate $k$NN queries on any given graph index, which lays the foundation of our hardness measure.
For graph indexes, the querying time cost is dominated by the time spent on distance calculations between the database points and the query point.
The number of Distance Calculations (NDC) is also the number of the accessed points in the graph when querying.
Thus, to analyze the query effort, we focus on calculating the minimum number of accessed points on a given graph (Section~\ref{sec:basic-me}).
This minimum effort (i.e., ME) provides a strict lower bound for the cost of any query answering algorithm.
However, since the common query algorithm is greedy, with a far higher cost than the theoretical bound, we pose three constraints when analyzing the ME to make it more practical (Sections~\ref{sec:cons-me} and ~\ref{me:decision-cost}).
Finally, to efficiently calculate the proposed ME (as well as our hardness measure), we map our definitions to the classical DST problems, and adapt DST solvers to our problem (Sections~\ref{sec:dst} and ~\ref{sec:alg-me}). 

\subsection{Basic ME Definition}
\label{sec:basic-me}
We first present a basic definition for ME.
Intuitively, it is defined as the minimum number of accessed points required to reach a given recall $Acc$ on a graph $G$.

\begin{definition}[$ME@Acc$]
\label{def:basic}
Consider a directed graph index $G(V,E)$, where $V$ is formed by the vectors in $db$, a query $q$ and its $k$NN set $N_k$.
Assume in the query process, one can only travel the graph along the edges $E$.
Then given a least recall requirement $Acc\in[0,1]$, $ME@Acc$ is defined as the minimum number of accessed vertices to obtain at least $Acc * k$ vertices in $N_k$, from any possible entry point $ep\in V$. 
\end{definition}

Figure~\ref{fig:me-illus} (a) and (b) show examples for $ME@Acc$ when $Acc$= 1 and 0.8 respectively.
When $Acc$=1, the optimal entry point is $v_1$ and it needs to access at least 12 points in $G$.
When $Acc$=0.8, the optimal entry point is changed to be $v_2$ and it only needs to obtain $0.8*5=4$ points in $N_k$ to reach 80\% recall@5.

\subject{Limitations of the basic ME definition.}
To achieve $ME@Acc$, three tough conditions should be satisfied:

\ding{172} \emph{Optimum entry point.} 
Every point $v\in V$ can be the entry point, and has a corresponding shortest routing path $p_v$\footnote{Actually, the accessed vertices and edges form a tree instead of a simple path as we allow backtracking. 
The shortest routing path means a minimum number of nodes in the tree. For simplicity, we still use the term ``path'' in the following.}. 
$ME$ is computed by starting from the entry point $v^*$ with the smallest $p_{v^*}$ among all points $v\in V$. $v^*$ is called the optimal entry point.

\ding{173} \emph{Infinite range of accessible candidates.} 
All neighbors of a visited point can be accessed without qualification, as long as they can provide shortcuts to the $k$NN.
Note that the greedy search algorithm requires that only vertices close enough to the query are qualified to be accessed, as described in Section~\ref{sec:graph-intro}.

\ding{174} \emph{Optimum search path and terminals.} 
Given the optimal entry point and the unlimited candidates that can be accessed, $ME$ requires finding the optimum search path.
Moreover, since we can select any $Acc * k$ points in $N_k$ as terminals, we should meanwhile select the optimum group of terminals that can be accessed with the shortest routing path.

Although these conditions are very strict, $ME@Acc$ is the tightest lower bound of the effort to answer a query on the given graph, for \emph{any} possible query algorithms, present and future.
Nevertheless, for common greedy search algorithms (Algorithm~\ref{alg:query}), this bound is inaccurate to represent the practical effort since these conditions are not satisfied.
Therefore, in the following section, we adjust $ME$ to the greedy search algorithm by adding three constraints.

\subsection{Adapt ME for Greedy Search}
\label{sec:cons-me}

To render the estimated ME closer to the actual effort of the current greedy search algorithm,
we constrain Definition~\ref{def:basic} in three ways: 
(1) the entry point should be selected from a limited range, 
(2) the range of accessible points in the search path should be limited, and 
(3) the optimal search path should satisfy the above two constraints.
More formally:
\begin{definition}[$ME_{\delta}^p@Acc$]
\label{def:me}
Consider a directed graph index $G(V,E)$, a query $q$ and its $k$NN set $N_k$.
Given a least recall requirement $Acc\in[0,1]$, a probabilistic lower bound $p \in (0,1]$, and a distance threshold $\delta$,
the constrained ME, denoted by $ME_\delta^p@Acc$, is defined as the number of points in a subgraph $Y(V_Y, E_Y)$, where $Y$ is induced from $G$ by $V_Y$, satisfying:

\ding{172} (Entry point) There exists a set $N_k' \subset N_k$ with $|N_k'| \geq p*k$.
For $\forall n \in N_k'$, there exists a path on subgraph $Y$ from $n$ to at least $Acc * k$ points in $N_k$.

\ding{173} (Range of accessible candidates) $\forall v \in V_Y$, $D(v,q)\leq (1+\delta)*d_k$.

\ding{174} (Optimum) Among all the qualified subgraph, $Y$ has the minimum number of vertices.

For clarity, such a subgraph $Y$ is also denoted by $Y^{\delta}_{p}@Acc$.

\end{definition}

In the following, we write $ME$ for short when it does not cause ambiguity.
Note that $Y^{\delta}_{p}@Acc$ does not always exist and in this case we assume $ME_\delta^p@Acc=\infty$.
We explain the rationale of this definition in a progressive way as shown in Figure~\ref{fig:me-illus}(b) to (e).

\begin{figure}[tb]
\includegraphics[width=.6\linewidth]{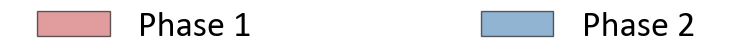}

\subfigure[HNSW]{
  \includegraphics[width=0.47\linewidth]{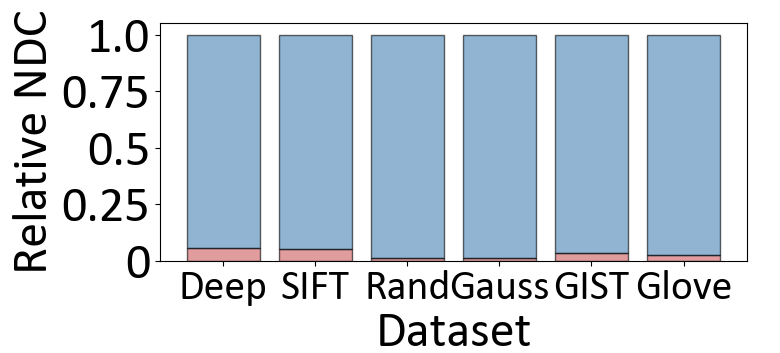}
}
\subfigure[NSG]{
\includegraphics[width=0.47\linewidth]{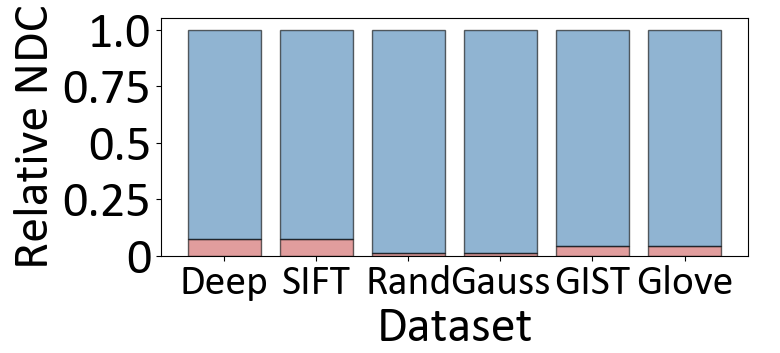}
}
\caption{Query time breakdown on 1,000 queries ($k$=50).}
\label{fig:bottleneck} 
\end{figure}

\subject{\ding{172} The entry point is constrained to be one of the $k$NN answers of the query.}
In the basic ME definition, we require a globally optimal entry point, from which we can construct the shortest search path.
However, it is very hard for the current algorithm to find such an entry point.
Nonetheless, we observe that using the current graph indexes, it is easy to obtain one of the $k$NN.
Specifically, we break the query process into two phases: Phase 1 starts from the entry point and continues until we first access one of the $k$NN points; Phase 2 starts immediately afterwards, and ends at the termination of the algorithm, as in~\cite{k-regular,note}.
As shown in Figure~\ref{fig:bottleneck}, Phase 1 costs less than 7\% of the total query time.
This indicates that in practice \emph{the bottleneck of the query process is Phase 2, not Phase 1}.
To this effect, we focus on modeling the effort in Phase 2, we omit the effort in Phase 1, and we use one of $k$NN as the entry point.
The example is shown in Figure~\ref{fig:me-illus}(c), where $n_2$ is selected as the optimum entry point which can reach at least 4 NNs in $Y$.
This change of the entry point only increases $ME$ by 1.

Since which nearest neighbor can be first accessed in Phase 1 is unknown and somewhat random, we further constrain $ME$ by allowing at least $p*k$ points in $N_k$ (i.e., points in $N_k'$) are able to reach $Acc*k$ points in $N_k$.
That is, as long as Phase 1 accesses one of the points in $N_k'$, the query cost in Phase 2 is bounded by $ME=|Y^{\infty}_{p@Acc}|$, where the symbol $\infty$ means there is no constraint on the range of accessible candidates of $Y$.
In this case, we view the parameter $p$ as a probabilistic lower bound to reach the target recall.
As shown in Figure~\ref{fig:me-illus}(d) where $p$ is set to be $40\%$, there are $40\%*5$ = 2 points who can reach at least 4 NNs.
Specifically, starting from $n_2$ we can reach $\{n_2,n_4,n_3,n_1\}$, and from $n_5$ we can reach $\{n_5,n_4,n_1,n_3\}$.
Here $N_k'=\{n_2,n_5\}$.

\subject{\ding{173} The accessible points in the search path are constrained to be near to the query.}
In our basic $ME$ definition, all the points in the graph are available to be accessed, as long as they can contribute to a shorter search path.
However, recall that in the current greedy search algorithm, the points that are closer to the query will be visited first, and the points that are farther than all the points in the $ef$-sized heap $H$ to the query will be skipped.
This greedy procedure limits the range of accessible points in the search path, i.e., no farther than the largest distance between the query and the points in $H$.
Consequently, we add a parameter $\delta$ to limit the accessible points when searching within $(1+\delta)*d_k$.
Notice that with a limited distance threshold, some shortcuts that contain distant points will be given up.
As shown in Figure~\ref{fig:me-illus}(e), $v_3$ is replaced by two closer points while $v_4$ and $v_5$ are replaced by four closer points, which increase $ME$ by 3.

\subject{Determine $\delta$ with the critical point $\delta_0$.}
In the greedy search algorithm, the distance threshold is determined on-the-fly by the heap $H$, which makes it impossible to model the exact case when analyzing the effort.
Nevertheless, we observe that there is a special value for parameter $\delta$, which is the minimum value such that there exists one qualified subgraph $Y$.
We name this value the \emph{critical point}, denoted by $\delta_0$.
Formally:
\begin{definition}[critical point $\delta_0$]
Given a least recall requirement $Acc\in[0,1]$, a probabilistic lower bound $p \in (0,1]$, the critical point $\delta_0^p@Acc$ is defined as the minimum distance threshold under which there exists a subgraph $Y^{\delta_0}_{p}@Acc$ satisfying the three constraints in Definition~\ref{def:me}.
\end{definition}
Note that $\delta_0$ describes the minimum requirement of the range of accessible candidates.
As $\delta$ increases from $\delta_0$, $ME_\delta^p@Acc$ is monotonically non-increasing and gradually converges to $ME_{\infty}^p@Acc$.
In other words, $\delta_0$ is the parameter that maximizes $ME_\delta^p@Acc$ among all possible $\delta$.
In practice, the quality of the points in heap $H$ quickly increases to a stable range in Phase 1 on high-dimensional data~\cite{promise}.
This indicates that $\delta$ is rather small in Phase 2, on which our $ME$ focuses.
In this case, a smaller $\delta$ can better reflect the actual querying situation, which is verified in the experiments in Section~\ref{sec:expr-ablation}.
We elaborate on how to find $\delta_0$ in Section~\ref{sec:alg-delta}.

\subsection{Incorporate Decision Cost into ME}
\label{me:decision-cost}
Besides the above constraints, there is still a major difference between $ME_\delta^p@Acc$ and actual query cost, i.e., the decision cost of each step when querying.
Remind that in Algorithm~\ref{alg:query}, to decide which point as the next stop, we need to compute the distance between each neighbor of the current point and the query.
These neighbors accessed in the greedy search algorithm are not included in $Y^{\delta}_{p}@Acc$ since they do not contribute to the navigability of $Y$, and thus out of the consideration of $ME_\delta^p@Acc$.
To this effect, we further modify our definition of ME to reflect the decision cost.

\begin{definition}[Decision cost]
\label{def:cost}
    Given a graph index $G(V,E)$, and a subgraph $Y$ of $G$, the decision set of a point $v\in Y$ is defined as
    \begin{equation}
        DS(v) = 
        \begin{cases}
        \{v\} & \text{if } v \text{has no out-neighbors in } Y
        \\
        \{x|x=v\vee(v,x)\in E\} & \text{otherwise}
        \end{cases}
    \end{equation}
    The decision cost of $v$ is the cardinality of the decision set, i.e. $Cost(v)=|DS(v)|$.
    Furthermore, the decision cost of the subgraph $Y$ is defined as $Cost(Y)=|\bigcup_{v\in Y}DS(v)|$.

\end{definition}

The decision cost of a point acts as the weight of the point, and then the definition of $ME$ is further modified as the minimum weight of the qualified subgraph.

\begin{definition}[$ME_{\delta}^p@Acc-exhaustive$]
\label{def:exhaustive}
$ME_{\delta}^p@Acc-exhaustive$ is defined by modifying the third constraint of $ME_{\delta}^p@Acc$, i.e.,

\ding{174} (Optimum with decision cost) Among all the qualified subgraph, $Y$ has the minimum cost $Cost(Y)$.
\end{definition}

We use $ME-exhaustive$ for short when no ambiguity.
As shown in Figure~\ref{fig:me-illus}(f), the neighbors in the original graph $G$ is also counted in $ME-exhaustive$, e.g., $v_3$ and $v_4$, despite that they do not contribute to the reachability.
This poses a punishment on the vertices with a large out-degree which introduce higher decision costs during querying.
For example, although $v_5$ can directly build a bridge from $n_5$ to $n_4$, the large out-degree incurs a high decision cost and thus it is replaced by three other points for the optimum.

By now, we have built the analytical framework for the ME of query answering in graph indexes.
All our definitions of ME can be utilized to describe different query algorithms on graph indexes.
It's clear that on the same graph and settings, $ME@Acc \leq ME_{\delta}^p@Acc \leq ME_{\delta}^p@Acc-exhaustive$.
In the following, we utilize $ME_{\delta_0}^p@Acc-exhaustive$ to describe the effort of the greedy search algorithm.
We also note that our analytical framework is flexible by combining different constraints on the basic format of $ME$ (Definition~\ref{def:basic}) such that it can be adapted for bounding the effort of more advanced search algorithms in the future.

\subsection{Map ME Definitions to DST Problems}
\label{sec:dst}
The core problem of solving $ME$ is to find a qualified subgraph $Y$, which is however, an NP-hard problem~\cite{dst}.
To calculate it efficiently, in this subsection, we prove $Y$ can be mapped to DST~\cite{dst} and its variants.
Specifically, we prove the equivalence or relevance between the DST problems and three definitions of ME, i.e., $ME@Acc$, $ME_{\delta}^p@Acc$ and $ME_{\delta}^p@Acc-exhaustive$ using Theorem~\ref{the:dst}, ~\ref{the:dsn} and ~\ref{the:exhaustive}, respectively.
Based on these analyses, we can calculate our $ME$ with the help of solutions for the classical DST problems. 

\subject{Map $\boldsymbol{ME@Acc}$ to DST.}
We start from $ME@Acc$ (Definition~\ref{def:basic}) where $Acc$=100\%.

\begin{definition}[Directed Steiner Tree (DST)]
    Given a directed graph $G(V,E)$, a root vertex $r\in V$, a group of terminals $T\subset V$, DST is the subgraph of $G$ with minimum edge cost satisfying that for $\forall v \in T$, there is a path from $r$ to $v$ on DST, denoted by $DST_T^r$ and the cost is denoted by $|DST_T^r|$.
\end{definition}

Note that the original DST problem is defined on edge-weighted graphs~\cite{dst}, and in this paper, we only study the unweighted version, which is a special case of the original problem.

\begin{theorem}
\label{the:dst}
Given a graph $G(V,E)$, a query $q$ with its $k$NN $N_k$, 
\begin{equation}
    ME@100\% = min_{r\in V}|DST_{N_k}^r|
\end{equation}

\end{theorem}

\subject{Map $\boldsymbol{ME_{\delta}^p@Acc}$ to vDSN.}
The constraint on the limited range of accessible candidates (\ding{173}) is equivalent to the minimum DST on a subgraph, while the limitation on the entry point in $N_k$ is equivalent to the minimum DST when the root $r\in N_k$.
We then use directed Steiner network (DSN) problem~\cite{dsn} to connect $ME$ with accuracy $Acc$ and probabilistic lower bound $p$ constraints.

\begin{definition}[vertex-focused Directed Steiner Network (vDSN)]
    Given a directed graph $G(V,E)$, a collection vertex pairs $P\subset V\times V$, vDSN is the subgraph of $G$ with minimum number of vertices satisfying that for $\forall (v,w) \in P$, there is a path from $v$ to $w$ on vDSN, denoted by $vDSN_P$ and the cost is denoted by $|vDSN_P|$.
\end{definition}

\begin{theorem}
\label{the:dsn}
    Given a directed graph $G(V,E)$, a query $q$ with its $k$NN $N_k$, a least recall $Acc$ and a probabilistic lower bound $p$, then
\begin{equation}
    ME_{\infty}^p@Acc = min|vDSN_P|       
\end{equation}
where $P\subset N_k\times N_k$, and the number of starting points in $P$ are at least $p*k$ and for each starting point, the number of different terminals are at least $Acc*k$.

\end{theorem}

\subject{Map $\boldsymbol{ME_{\delta}^p@Acc-exhaustive}$ to node-weighted vDSN.}
For $ME-exhaustive$, there is no specific Steiner-related problem matching the definition.
However, we can upper bound it by node-weighted vDSN problem, that is, the target is modified to find the minimum sum weights of vertices in the subgraph.
For simplicity, we still denote the cost by $|vDSN_P|$.

\begin{theorem}
\label{the:exhaustive}
    Given a graph $G(V,E)$, a vertex weighting function $W(v)=Cost(v)$ for $v\in V$, we have
\begin{equation}
    ME_{\infty}^p@Acc-exhaustive \leq min|vDSN_P|       
\end{equation}
where $P$ has the same requirements as Theorem~\ref{the:dsn}.

\end{theorem}

\subsection{Calculate ME using Steiner Tree Solvers}
\label{sec:alg-me}
Based on these theorems, the three definitions of ME is mapped to the DST, vDSN, and node-weighted vDSN problems.

Finding the exact DST is an NP-hard and APX-hard problem~\cite{steiner1,dsn-99}, which means that there is even no polynomial-time constant-approximation algorithm.
To the best of our knowledge, the exact~\cite{ilp,thesis} and approximate~\cite{steiner1,dsn-99,dst-14} algorithms are very time-consuming.
Meanwhile, efficient heuristic algorithms can provide high-precision results in practice~\cite{thesis,pace,dst-repo}.
Therefore, we use the shortest path-based algorithm to find the shortest paths from root vertex $r$ to each of the terminals $t\in T$ and unite them to build DST.

The DSN problem is harder than DST problem.
None of exact~\cite{dsn-exact} algorithms or approximate~\cite{dsn-20,dsn-17,dsn-12,dsn-99} algorithms are efficient.
To this effect, we design a heuristic algorithm.
We first group the set $P$ of vertex pairs by the starting vertex and reduce the DSN problem to multiple DST problems.
Then we find DSTs one by one as described above and unite them to form the final result.

Lastly, to efficiently estimate $ME-exhaustive$, we adapt the shortest path-based algorithm with the node-weighted Dijkstra algorithm~\cite{dijkstra}.
The weight of a node is set to the cost defined in Definition~\ref{def:cost}.
We omit the algorithm details due to the lack of space.

\section{Find the Critical Point $\delta_0$}
\label{sec:alg-delta}
Given the solutions for ME, we still need algorithms to calculate the critical point $\delta_0$, and the optimum set of vertex pairs $P$.
In this subsection, we propose an efficient exact algorithm to achieve these targets at the same time.

\subject{The basic structure of the algorithm.}
To obtain $\delta_0$, the basic algorithm is as follows. First, we initialize $Y$ as empty and then add vertices to $Y$ in the ascending order of distance to the query one by one. After each insertion, we check whether the current subgraph $Y$ is a qualified $Y_{p}^\delta@Acc$.
In this process, the list of the sorted vertices can be supported by many 
exact $k$NN indexes in CPU~\cite{exact-cpu,exact-cpu2,faiss} and algorithms in GPU~\cite{exact1,exact2,exact3,faiss}, and the major time cost comes from the graph qualification.
To check the graph, a naive method is to 
iterate all the points in $N_k$ and for each point, we use BFS to obtain their reachable points in $N_k$.
Once the number of points that can reach at least $Acc*k$ points in $N_k$ on $Y$ is no less than $p*k$, the current graph is qualified and $\delta_0 = D(n_i,q)$, where $n_i$ is the latest inserted point.
Obviously, the repeated reachability checking is redundant and incurs unnecessary costs.

\subject{Avoid redundant computation by maintaining reachability information on the fly.}
we design a novel algorithm to maintain the reachability information of all inserted points on the fly.
The basic idea is to group together the points that are reachable to each other in $Y$ as a union-find set, (i.e., the reachable group), and maintain the reachability between different reachable groups through a graph (i.e., union-find set graph (USG)).
When we check the reachability of a point, we can directly apply BFS on the small USG instead of the complete subgraph $Y$ to reduce the complexity.

\begin{figure}[tb]
  \includegraphics[width=\linewidth]{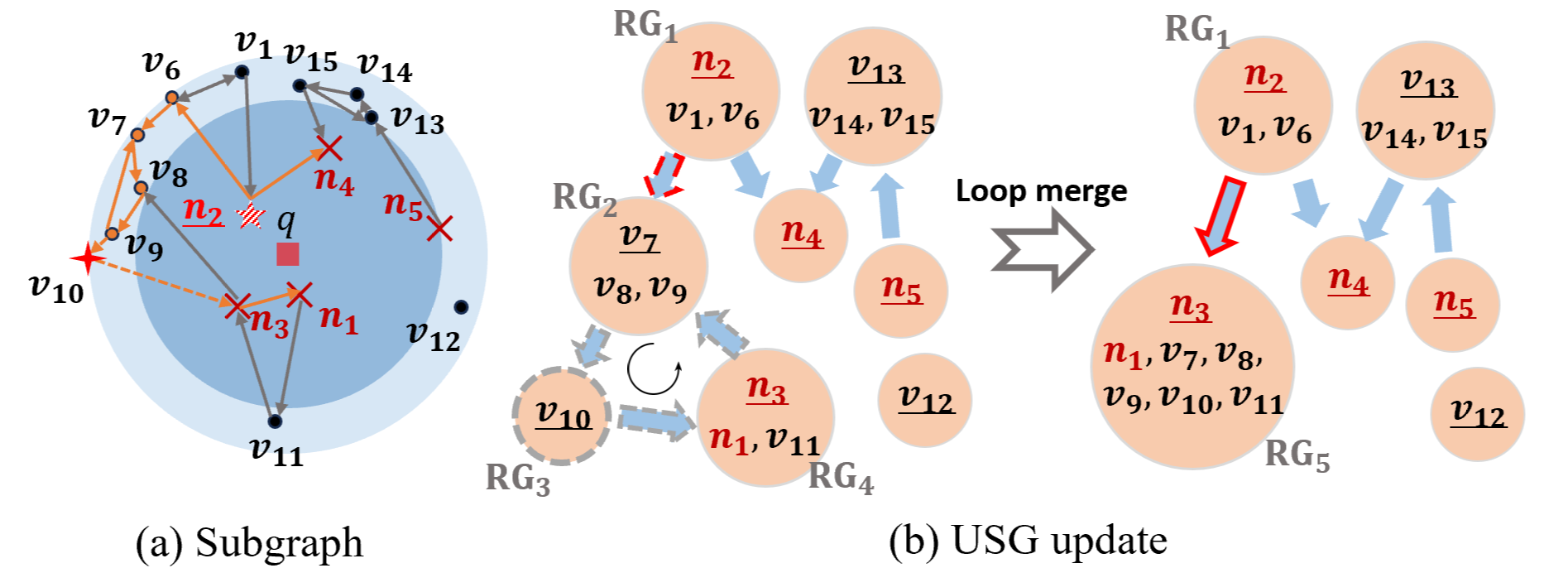}
\vspace{-\baselineskip}
  \caption{The updates of USG when adding a new vertex $v_{10}$ into the subgraph. In (b), the vertices in the same circle form a reachable group.}
  \label{fig:alg}
\end{figure}

\subject{Update USG by loop detection.}
To minimize the size of USG and thus the complexity of the reachability check, we merge all the reachable groups belonging to a loop in USG as a single one.
That is, after updating, USG will be a directed acyclic graph (DAG).
See Figure~\ref{fig:alg} as an example, where $v_{10}$ is newly inserted.
We first add $v_{10}$ as an individual reachable group into USG, along with corresponding edges.
After that, we detect a new loop in USG that contains three reachable groups (i.e., $RG_1$, $RG_2$, and $RG_3$). 
Then we merge these three reachable groups into a single one in USG (i.e., $RG_5$), and also unite their corresponding union-find sets.
Note that the edges inside the loop will be removed while the edges that connect reachable groups outside the loop will be reserved for the new reachable group (e.g., the edge connecting $RG_1$).
In this way, the subgraph with 16 vertices and 19 edges is reduced to the USG with 6 vertices (i.e., reachable groups) and 4 edges.
The following theorems prove that the loop merge can keep the property of a reachable group.
The proof is omitted due to lack of space.


\begin{theorem}
\label{theorem:reachable-groups}
    Given a subgraph $G_i(N_i,E_i)$, two reachable groups $RG_1, RG_2 \subset N_i$, $RG_1\cap RG_2 = \varnothing$, then if $\exists v_1 \in RG_1, v_2 \in RG_2$, and $v_1$ can reach $v_2$ in $G_i$, then $\forall v \in RG_1, w \in RG_2$, $v$ can reach $w$ in $G_i$.
\end{theorem}


\begin{corollary}
\label{corollary:reachable-check}
    In a USG of a subgraph $G_i$, if there exists a loop, $RG_i\rightarrow RG_{i+1} \rightarrow \dots RG_j \rightarrow RG_i$, then for $\forall v,w \in RG_t, i\leq t \leq j$, $v$ can reach $w$ in $G_i$.
\end{corollary}

In the implementation, we use the classical Depth-First-Search (DFS) algorithm to detect whether there is a loop on USG.
This loop detection is executed every time a new point is inserted into USG.
After detecting the loop, 
we collect all the incoming and outcoming edges that point to and start from the reachable groups in the loop, respectively.
Then we remove all these reachable groups and related edges from USG, and insert a new vertex into USG, along with the collected edges beforehand for this new vertex.

The details of the algorithm are shown in Algorithm~\ref{alg:delta0}.
After preparing the NN list and initialization (lines 1-3), the main body of the algorithm is a loop that takes one NN to increment the subgraph each time (lines 5-7).
The loop can be divided into three stages.
In the first stage (lines 8-12), we locate the in- and out-neighbors of the new point and add edges on USG.
In the second stage (lines 13-15), we detect and merge the loops in USG.
The union-find sets are also updated accordingly.
In the third stage (lines 16-23), we check the reachability on USG.
Then for each root, we use BFS to find which points in $N_k$ are reachable from the root and count qualified points (lines 19-25).
Finally, $D(n_i,q)$ is returned as $\delta_0$ along with the found vertex pairs set $P$ (line 27).

\begin{algorithm}[tb]
\caption{Find $\delta_0$ and $P$ (graph $G$, query $q$, reversed graph $revG$, $k$, recall target $Acc$, probabilistic lower bound $p$)} 
\label{alg:delta0}
{\scriptsize
\begin{algorithmic}[1]
\STATE Get sufficient number of $NN$ of $q$.
\STATE Initialize an empty union-find set $UF$, an empty union-find set graph $USG$, and an empty hash set $Ins$.
\STATE Initialize the result vertex pair set $P$.
\WHILE{True}
    \STATE $n_i$ = pop the first element from $NN$.
    \STATE Add $n_i$ as an individual union-find set to $UF$.
    \STATE Add $n_i$ as an individual vertex in $USG$.
\item[] \(\triangleright\) \textit{(1) Subgraph construction.}
    \STATE $N_{out}=G[n_i]\cap Ins$, $N_{in}=revG[n_i]\cap Ins$.
    \FOR{$n_{out}$ in $N_{out}$}
        \STATE Add an directed edge from $n_{i}$ to $UF[n_{out}]$ in $USG$, if not exists.
    \ENDFOR
    \FOR{$n_{in}$ in $N_{in}$}
        \STATE Add an directed edge from $UF[n_{in}]$ to $n_{i}$ in $USG$, if not exists.
    \ENDFOR
    \item[] \(\triangleright\) \textit{(2) Reduce $USG$ to a DAG by loop detection.}
    \WHILE{there exists a loop in $USG$}
        \STATE Unite all the $RG$ in this loop as a single $RG$.
        \STATE Unite all the union-find sets in this loop in $UF$.
    \ENDWHILE
    \item[] \(\triangleright\) \textit{(3) Qualification check: the reachability of $k$NN.}
    \STATE Initialize $R_k$ as an empty hash map, where key is the root of a union-find set, and the value is a set containing the elements in this union-find set.
    \FOR{$1\leq j\leq min(k,i)$}
        \STATE $R_k[UF[n_j]]$.add($n_j$)
    \ENDFOR
    \FOR{$(s, Set) \in R_k$}
        \STATE Use BFS to find which keys in $R_k$ are reachable in $USG$ from $s$, and store them into a set $reach$.
        \STATE $reach$ = $reach \cup Set$
        \IF{$|reach| \geq Acc*k$}
            \STATE $P$.add($(s,t)$), for $\forall t \in reach$
        \ENDIF
    \ENDFOR
    \IF{$|P.keys().distinct()| \geq p*k$}
        \STATE break
    \ENDIF
    \STATE Add $n_i$ into $Ins$ and clear $P$.
\ENDWHILE
\RETURN $D(n_i,q)$ as $\delta_0$, and $P$
\end{algorithmic}
} 
\end{algorithm}

\subject{Complexity Analysis.}
Assuming the average number of iterations is $\hat{it}$, the complexity of the first stage is $\mathcal{O}(\hat{it})$ since $\hat{it}$ will be the expected size of $Ins$, and also the number of vertices in the complete subgraph.
For the second stage, the loop detection costs $\mathcal{O}(|USG|)$ where $|USG|$ denotes the sum number of vertices and edges in USG, usually ranging from tens to a few hundred, while loop merge needs only constant time.
The third stage needs at most $k$ times of BFS search, so the complexity is $\mathcal{O}(k|USG|)$.
Overall, the time complexity is $\mathcal{O}(\hat{it}(\hat{it} + k|USG|))$.
In contrast, the complexity of the naive method is $\mathcal{O}(k\hat{M}\hat{it}^2)$, where $\hat{M}$ is the expected average out-degree of the subgraph, usually ranging from tens to hundreds.
Given that $\hat{it}$ usually ranges from hundreds to tens of thousands, our algorithm can provide $k*\hat{M
}$ times speedup according to this complexity analysis.

\section{$\boldsymbol{Steiner}$-Hardness}
\label{sec:hardness}
Now we describe our \emph{index-independent} $Steiner$-hardness for queries on all current graph indexes, based on our definitions of ME.
Specifically, we use $ME_{\delta_0}^p@Acc-exhaustive$ on a representative graph structure, (approximate) MRNG to define $Steiner$-hardness. 
\begin{definition}[$Steiner$-hardness]
    Given a dataset $db$ we build approximate $MRNG$ on $db$ with a parameter $efC$. 
    Then, $Steiner$-hardness for a query $q$ is defined as 
    \begin{equation}
        Steiner_q = ME_{\delta_0}^p@Acc-exhaustive \textit{ on MRNG} 
    \end{equation}
    where $Acc$ is the given recall requirement, and $p$ is the probabilistic lower bound, as 
    in Definition~\ref{def:exhaustive}.
\end{definition}

To build MRNG, for any point $v$, all other points in $db$ are sorted according to the distance to $v$ in ascent.
Then we insert edges from $v$ to these points one by one, as long as the angle between the new edge and all existing edges are no smaller than 60\textdegree.
Since all the current graph indexes only select neighbors from a pool of NNs of limited size $efC$~\footnote{It is denoted by $efC$ in HNSW, $C$ in NSG and $\tau$-MNG, and $k_{ext}$ in DEG.}, we approximate MRNG by also considering only $efC$ NNs rather than all points.
In this case, the approximate MRNG can be viewed as a pruned version of $K$Graph where $K=efC$.
The reasons for selecting approximate MRNG as the representative graph are as follows.

\begin{figure}[tb]
\subfigure[Deep]{
\includegraphics[width=0.47\linewidth]{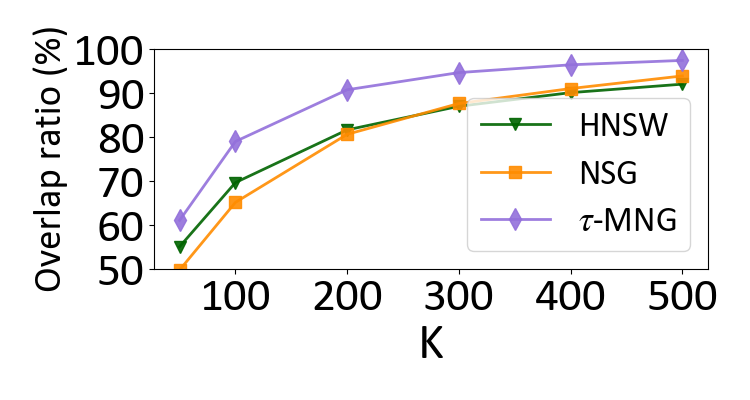}
}
\subfigure[Gauss]{
\includegraphics[width=0.47\linewidth]{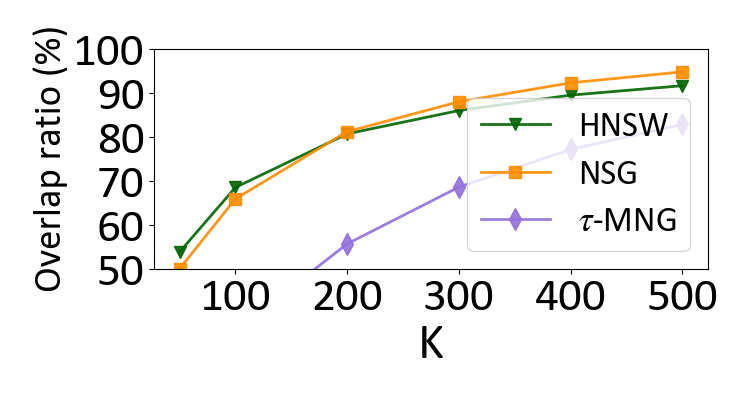}
}
\caption{Overlap ratio of edges in current graph indexes and $K$Graph with varying $K$.}
\label{fig:overlap} 
\end{figure}

(1) \emph{Approximate MRNG and current graph indexes are all pruned versions of a large $K$Graph.}
As shown in Figure~\ref{fig:overlap}, we count the percentage of the edges in graph indexes that occur in $K$Graph.
The results show that over 80\% of the edges of current graph indexes are in a $K$Graph with $K\geq 500$.
Therefore, approximate MRNG represents a common structure for advanced graph indexes.

(2) \emph{Both approximate MRNG and other graph indexes limit the number of edges.}
Approximate MRNG selects edges according to the spatial distribution with bounded out-degree.
Furthermore, it is proved to own a nice navigability~\cite{theory-understand}.
Thus, approximate MRNG reaches a balance of sparsity-navigability trade-off, which matches the rationale of current graph indexes.

(3) \emph{The structure of MRNG only relies on the data distribution in high-d space and is not influenced by randomness or any particular index-building strategy.}


\subsection{Unbiased Workload Generation}
\label{sec:workload-gen}
Based on our $Steiner$-hardness, we can build unbiased query workloads for evaluating graph indexes.
To comprehensively and fairly evaluate the graph index, the produced queries must follow the following two principles: 
\begin{enumerate}
    \item \textbf{Validity}. The queries should follow the distribution of the dataset.
    \item \textbf{Uniformness}. The hardness of the queries should follow a uniform distribution across the hardness spectrum.
\end{enumerate}

Validity avoids the queries which are out of the distribution of the dataset.
For example, in a dataset where the values in the vectors are between 0 and 1, a query vector with values beyond 100 is meaningless for evaluation.
Uniformness guarantees that the workload will not exhibit bias neither for simple,  nor for hard queries. 
Based on these two principles, the core idea of our approach is a generation-and-check method.
That is, we first generate sufficient vectors by over-sampling from the dataset~\footnote{In the query set of the original workload, the hard query is usually insufficient for evaluation. So we need to generate more queries as candidates.} and then check the hardness of these vectors. 
Finally we preserve the vectors that satisfy the uniformness principle.

Specifically, we use Gaussian Mixture Model (GMM)~\cite{gmm} to learn the data distribution of the dataset.
According to our pre-experiments, the distribution of the Mahalanobis distance~\cite{ma-dist} -- which is adopted to test whether two groups of vectors are in the same distribution~\cite{ood} -- on generated data is very similar to the dataset.
It verifies the effectiveness of GMM.
We use the GMM model to produce sufficient candidate vectors, and compute $Steiner$-hardness of them.
Secondly, we build histograms on the new data w.r.t. their $Steiner$-hardness.
Specifically, we remove the extreme values, i.e., too simple and hard queries, and then uniformly split the range of hardness by $h$ equi-length segments.
Given the cardinality of the new workload as $Q$, we randomly sample $\lceil Q/h\rceil$ queries from each segment to build the workload.
In this way, the new workload encompasses the entire spectrum of $Steiner$-hardness.




\begin{figure*}[tb]
\includegraphics[width=.45\linewidth]{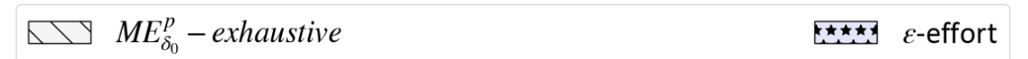}

\subfigure[GIST]{
\includegraphics[width=0.235\linewidth]{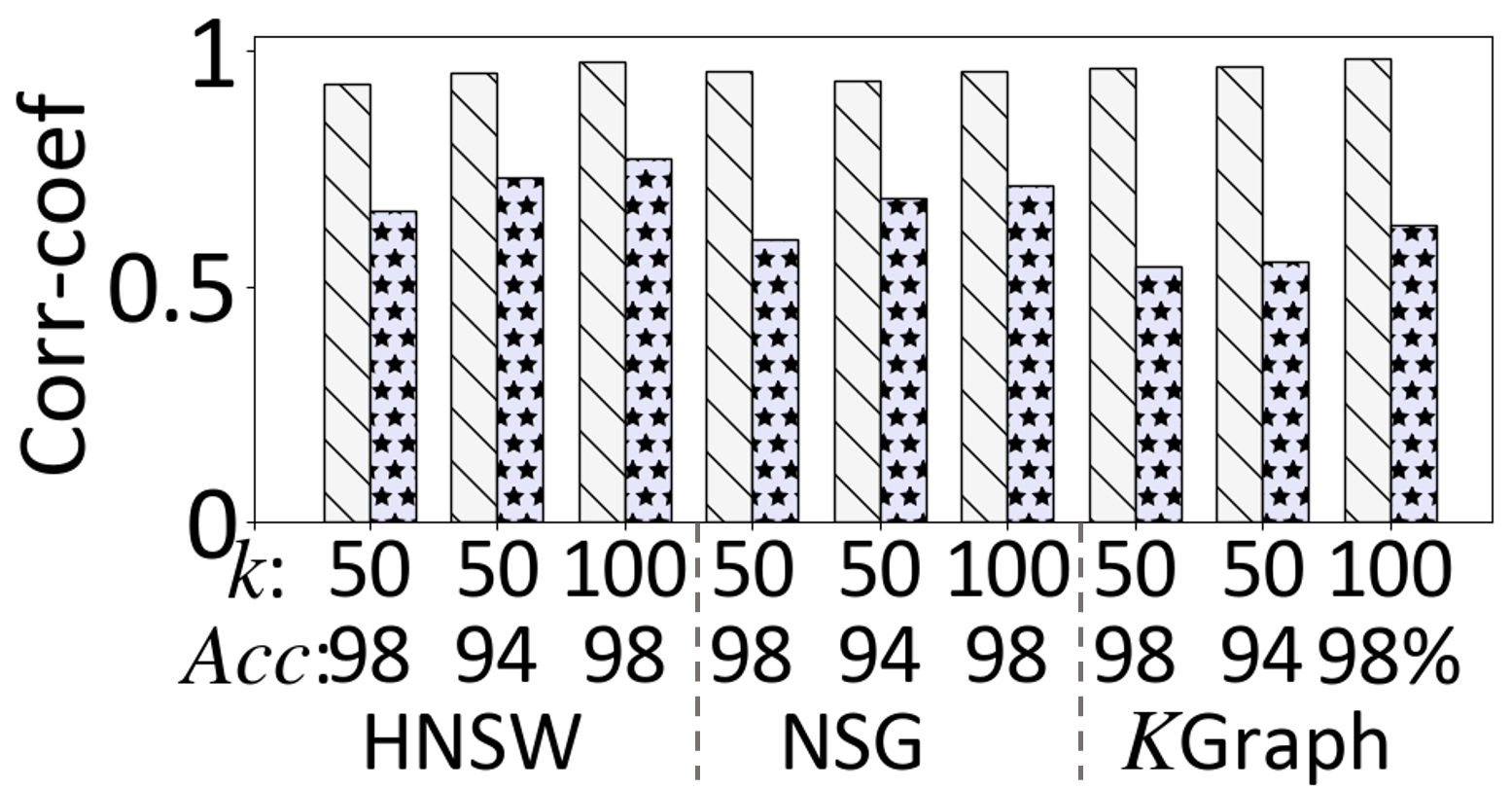}
}
\subfigure[Rand]{
\includegraphics[width=0.235\linewidth]{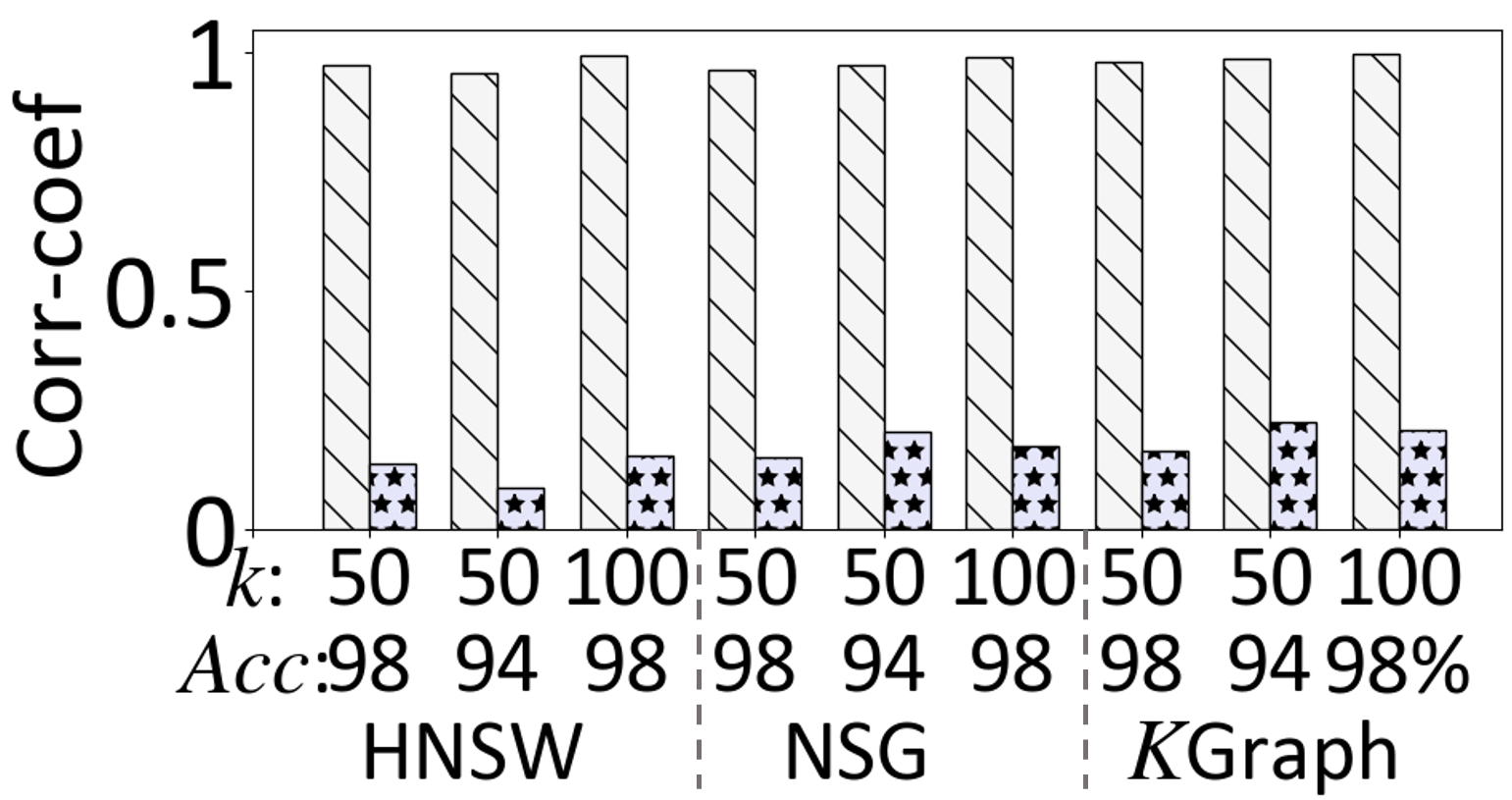}
}
\subfigure[Glove]{
\includegraphics[width=0.235\linewidth]{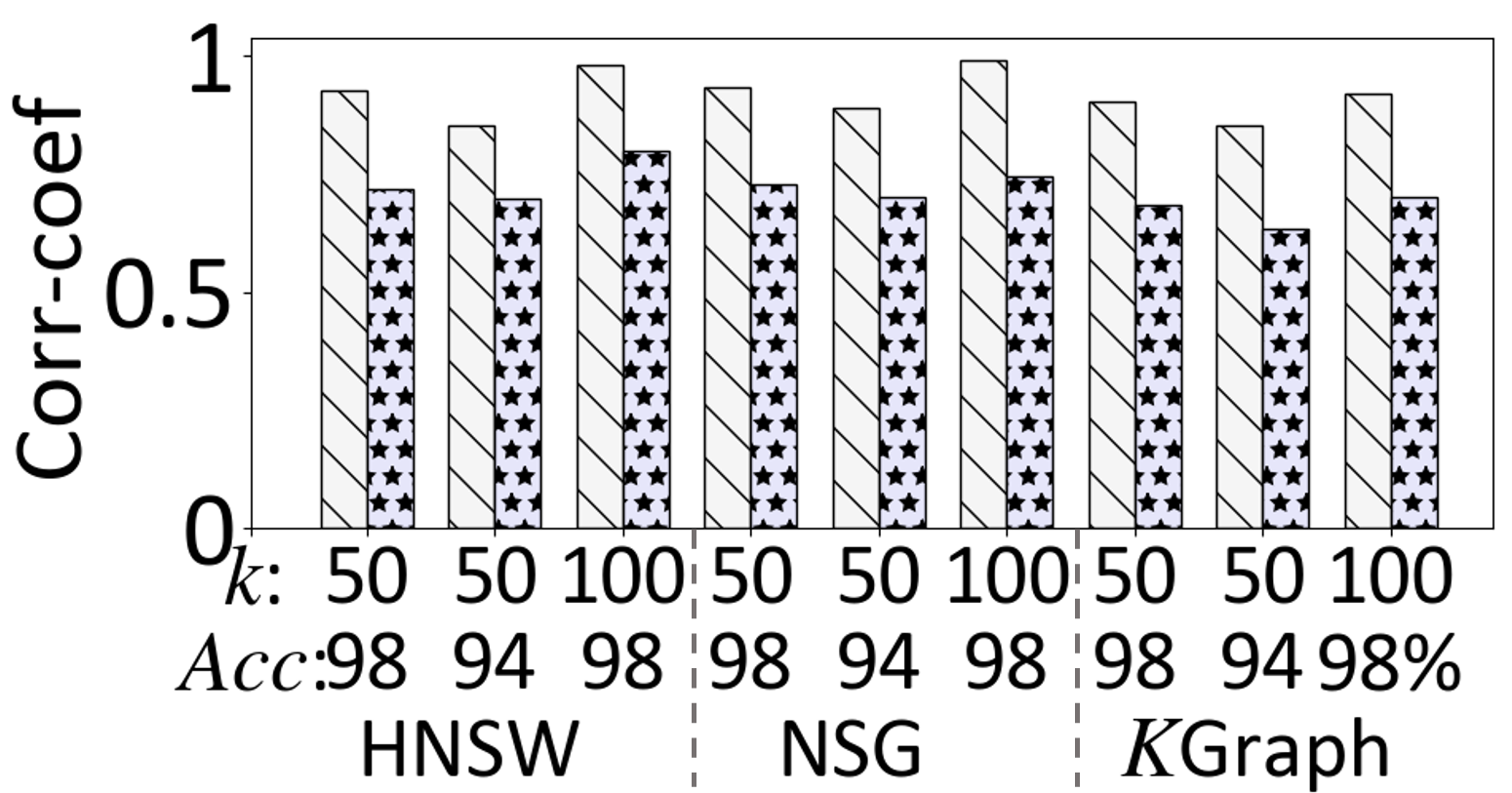}
}
\subfigure[Tiny10M]{
\includegraphics[width=0.235\linewidth]{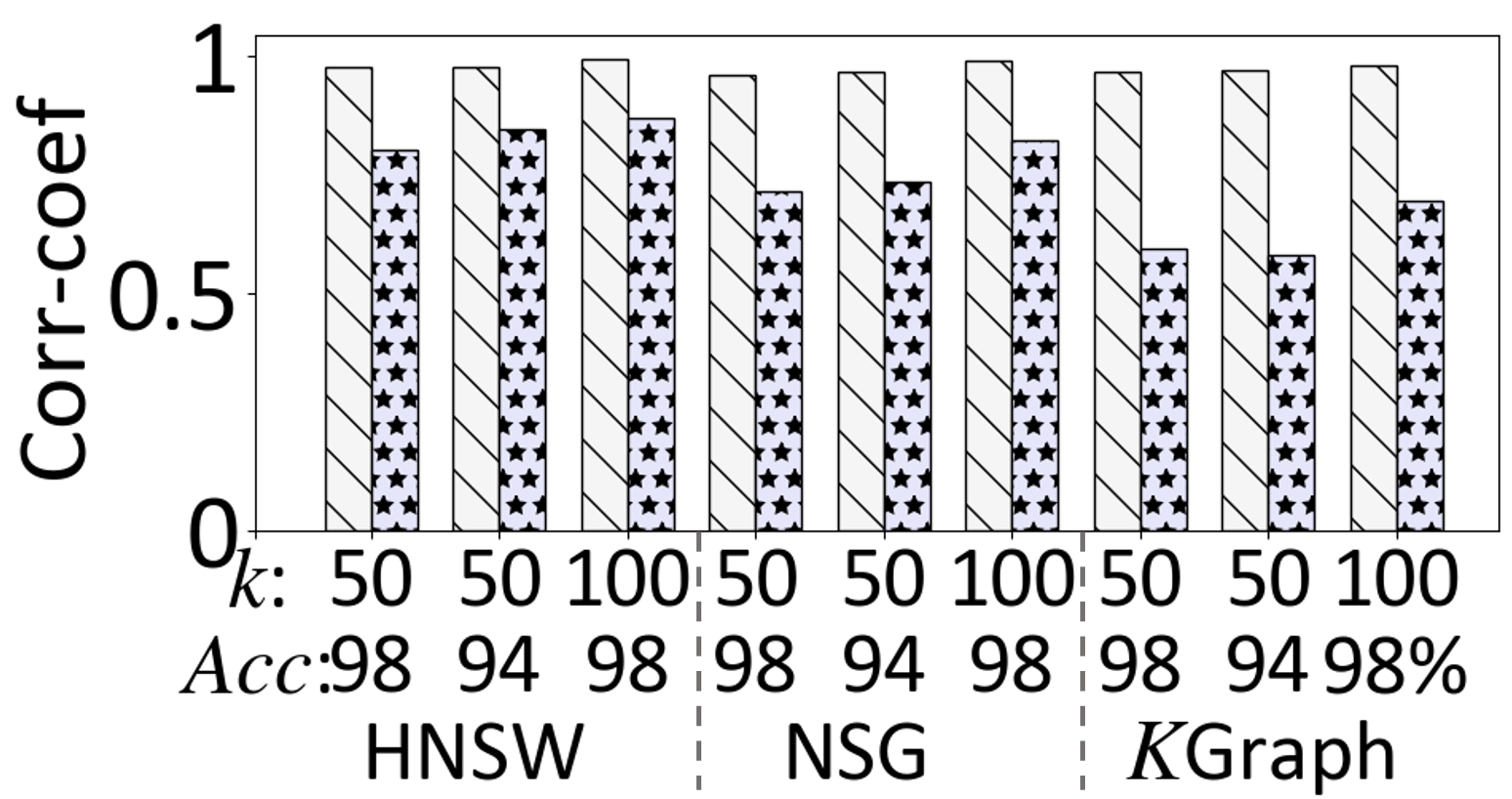}
}
\caption{Pearson correlation coefficients between $ME_{\delta_0}^p-exhaustive$ and actual query effort (NDC).}
\label{fig:me}
\end{figure*}

\begin{figure*}[tb]
\includegraphics[width=.85\linewidth]{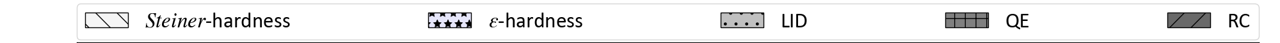}
\subfigure[Deep]{
\includegraphics[width=0.187\linewidth]{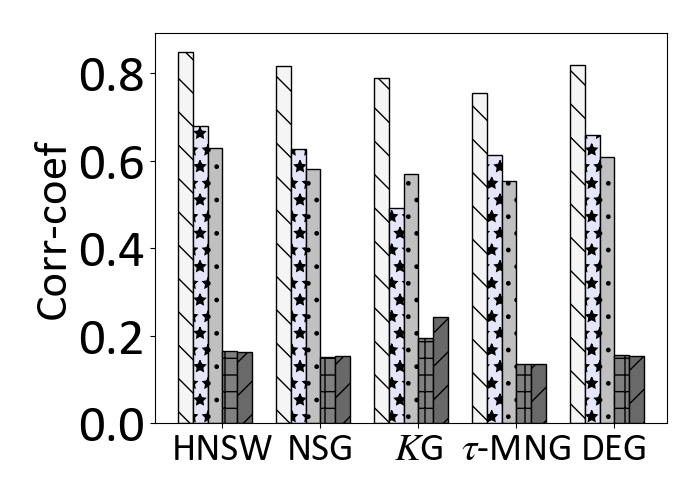}
}
\subfigure[GIST]{
\includegraphics[width=0.187\linewidth]{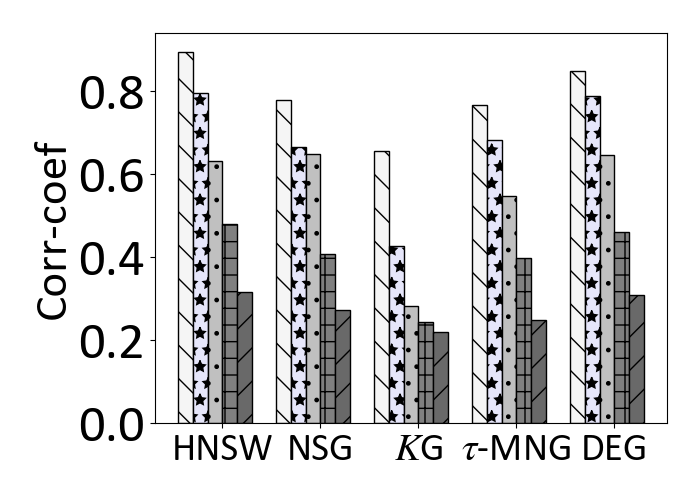}
}
\subfigure[Rand]{
\includegraphics[width=0.187\linewidth]{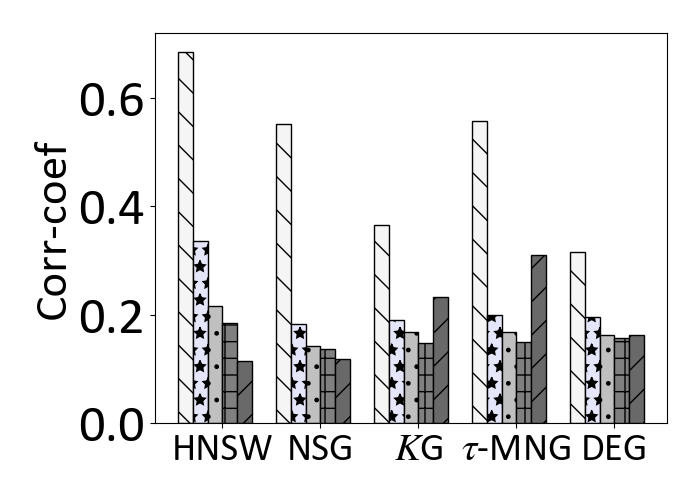}
}
\subfigure[Glove]{
\includegraphics[width=0.187\linewidth]{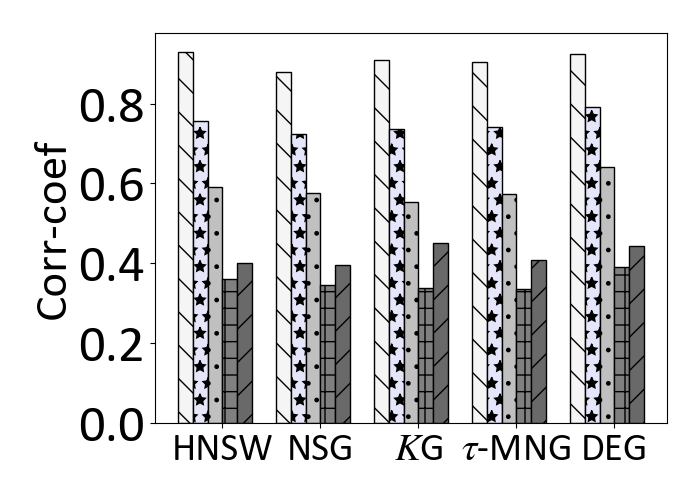}
}
\subfigure[Tiny10M]{
\includegraphics[width=0.187\linewidth]{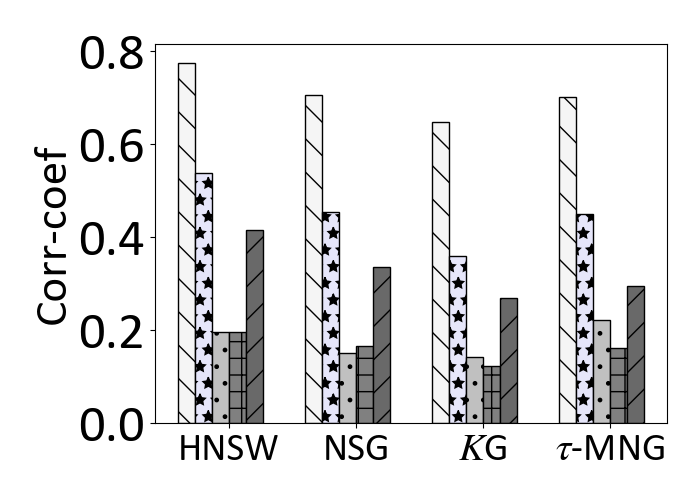}
}
\caption{Correlation coefficients between different hardness measures and the average query effort (NDC)}.
\label{fig:hardness}
\end{figure*}

\begin{figure}[tb]
\includegraphics[width=\linewidth]{hardness-legend.png}
\subfigure[Varying $Acc$]{
\includegraphics[width=0.475\linewidth]{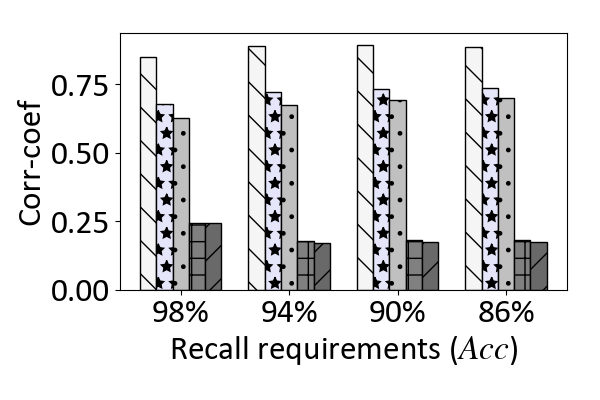}
}
\subfigure[Varying $k$]{
\includegraphics[width=0.475\linewidth]{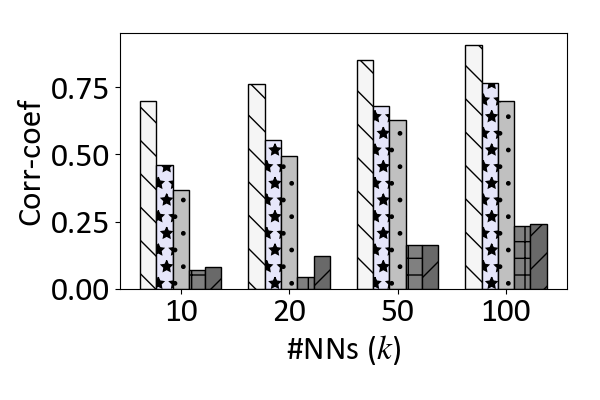}
}
\caption{Correlation coefficients of hardness measures under different parameters.}
\label{fig:hardness-para}
\end{figure}

\section{Experiments}
\label{sec:expr}
\subsection{Experimental Setup}
Experiments were conducted on an machine with two 40-core Intel(R) Xeon(R) Platinum 8383C 2.70GHz CPUs and 1TB DDR4 main memory.
All the codes are implemented in C/C++, compiled by g++ 10.1.0 with -O3 optimization. 
We use five public datasets and two synthetic dataset with varying sizes and dimensionalities, as shown in Table~\ref{tab:dataset}.
The results of Gauss dataset are omitted since they are similar to Rand dataset. 
Other omitted datasets are shown in our code repository.
Since only HNSW and $K$Graph can be built on Deep1B in our machine due to memory limitations, in this paper, we only use Deep1B to test the scalability of our approaches.


\subject{Algorithms and parameters.}
We select five representative graph indexes for evaluation, including four advanced indexes and the $K$Graph as a baseline.
HNSW~\cite{hnsw} and NSG~\cite{nsg} are the state-of-the-arts according to previous benchmarks~\cite{wang2022graphbased}.
We only test the base layer of HNSW for evaluation, since the upper layers of HNSW only provide benefits in very low dimensionality~\cite{bulletin-wang,promise}.
We select $\tau$-MNG~\cite{tau} and DEG~\cite{k-regular} as the latest indexes.
We disable the dynamic edge optimization of DEG to control the construction time as in the original paper.
The parameters for building these indexes are tuned for the best query performance.
When evaluating our analytical framework, $Acc$ and $k$ are set to 98\% and 50, respectively by default.
$efC$ is set to 2048.
The default dataset and index are Deep and HNSW.
We tune $p$ in the range 0.85\textasciitilde 1 to achieve the optimal correlation.

\begin{table}[!tb]
    \centering
    \footnotesize
    \caption{Dataset statistics.}
    \begin{tabular}{c|ccccc}
    \toprule
        ~ & Base size & Query size & Dim. & Source & Type\\ \hline
        Deep & 1,000,000 & 10,000 & 96 & \cite{deep-dataset} & Image \\ \hline
        GIST & 1,000,000 & 1,000 & 960 & \cite{sift-dataset} & Image\\ \hline
        Glove & 1,183,514 & 10,000 & 100 & \cite{glove}& Text \\ \hline
        Rand & 1,000,000 & 10,000 & 100 & $U(-1,1)$& Syn. \\ \hline
        Gauss & 1,000,000 & 10,000 & 100 & $N(0,1)$ & Syn.\\ \hline
        Tiny10M & 10,000,000 & 1,000 & 150 & \cite{hvs} & Image \\ \hline
        Deep1B & 1,000,000,000 & 10,000 & 96 & \cite{deep1b-dataset} & Image \\ \hline
    \end{tabular}
    \label{tab:dataset}
\end{table}

\subsection{Effectiveness of ME}
We first evaluate whether $ME_{\delta_0}^p@Acc-exhaustive$
can describe the actual query effort on the graph indexes.
Given a query and a recall target, we use the least NDC to reach the recall as the query effort, 
As shown in Figure~\ref{fig:me}, the correlation coefficients of $\epsilon$-effort are on average \textbf{0.66}, \textbf{0.17}, \textbf{0.70}, and \textbf{0.74} in GIST, Rand, Glove and Tiny10M datasets for all evaluated indexes\footnote{$\tau$-MNG and DEG indexes and Deep dataset are omitted due to lack of space, whose results are similar to other settings.}, whereas our $ME-exhaustive$ is \textbf{0.95}, \textbf{0.98}, \textbf{0.92} and \textbf{0.98} respectively.
Clearly, $ME-exhaustive$ can describe the query effort more precisely than $\epsilon$-hardness due to the focus on the graph connections.

\begin{figure*}[tb]
\includegraphics[width=0.475\linewidth]{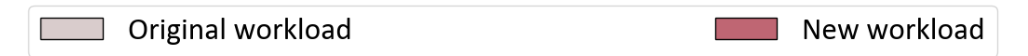} 

\begin{minipage}{0.475\linewidth}
\subfigure[Deep]{
\includegraphics[width=0.31\linewidth]{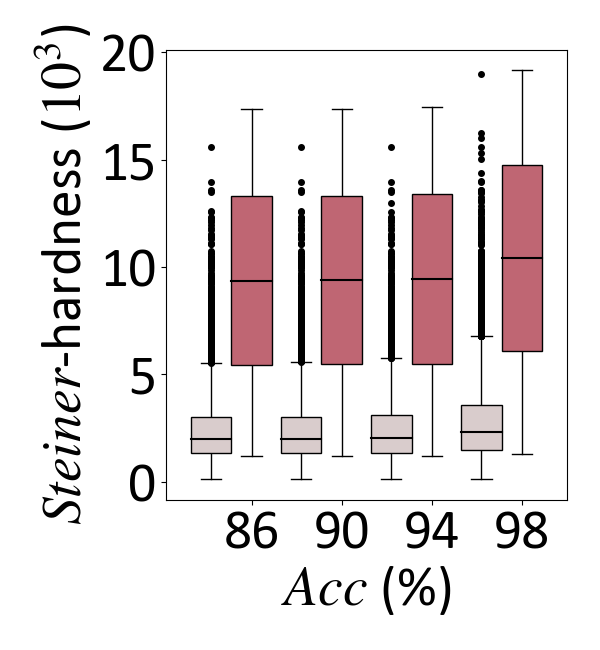}
}
\subfigure[Glove]{
\includegraphics[width=0.31\linewidth]{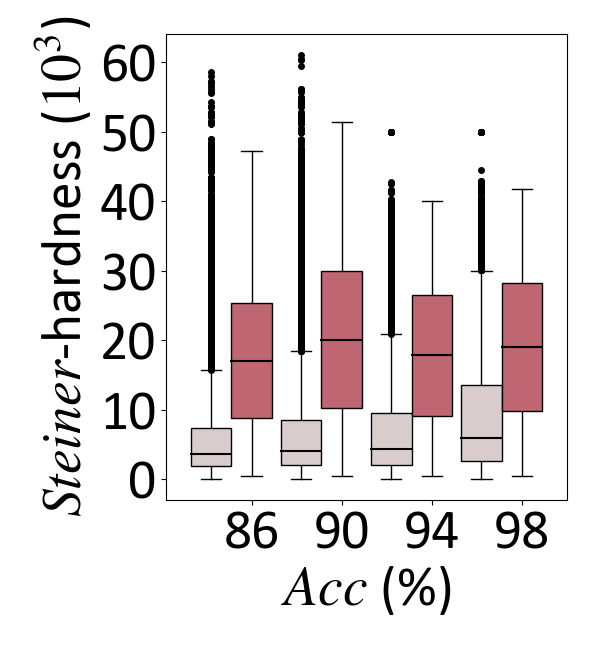}
}
\subfigure[Tiny10M]{
\includegraphics[width=0.31\linewidth]{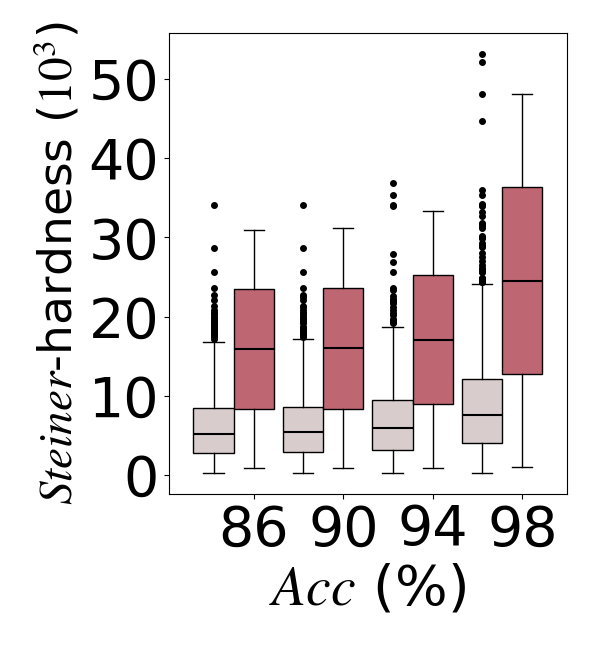}
}
\caption{Distribution of query hardness in the workloads.}
\label{fig:dist} 
\end{minipage}
\begin{minipage}{0.475\linewidth}
\subfigure[Deep]{
\includegraphics[width=0.31\linewidth]{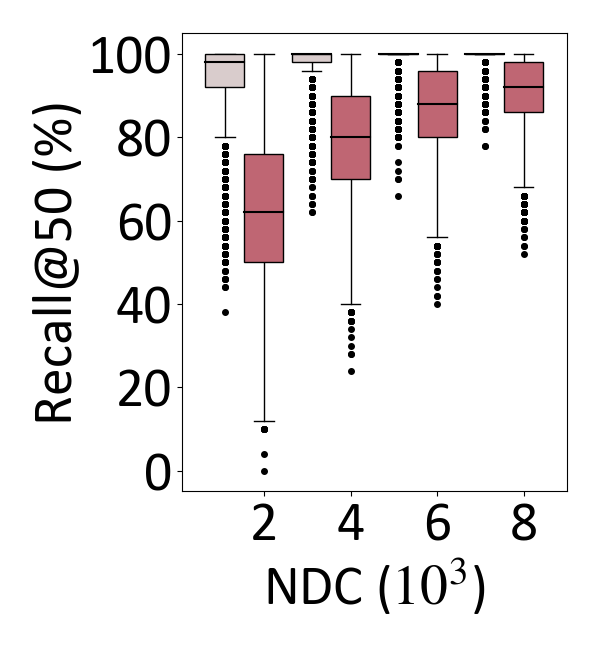}
}
\subfigure[Glove]{
\includegraphics[width=0.31\linewidth]{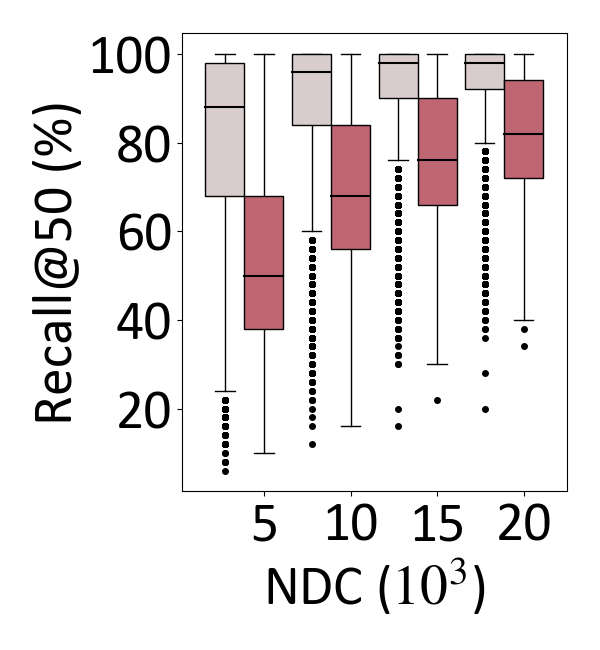}
}
\subfigure[Tiny10M]{
\includegraphics[width=0.31\linewidth]{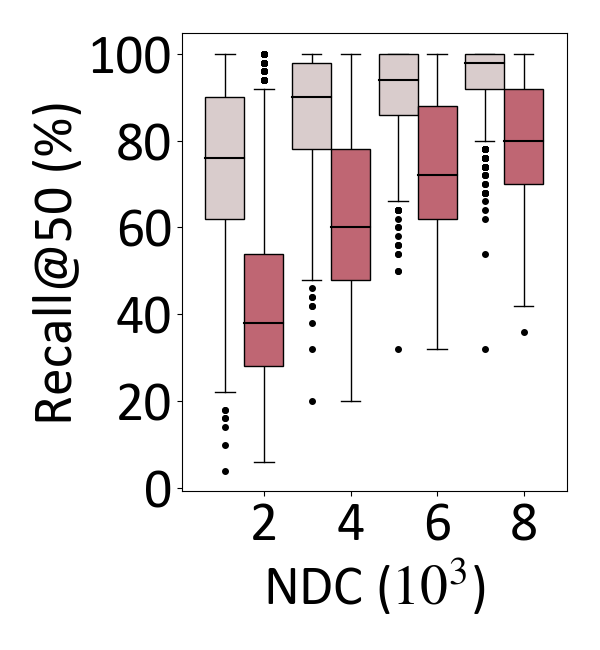}
}
\caption{Distribution of recall in the workloads.}
\label{fig:recall-diff}
\end{minipage}
\end{figure*}

\begin{figure*}[tb]
\includegraphics[width=.9\linewidth]{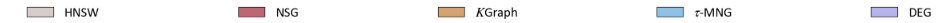}

\begin{minipage}{0.475\linewidth}
\subfigure[Deep]{
\includegraphics[width=0.31\linewidth]{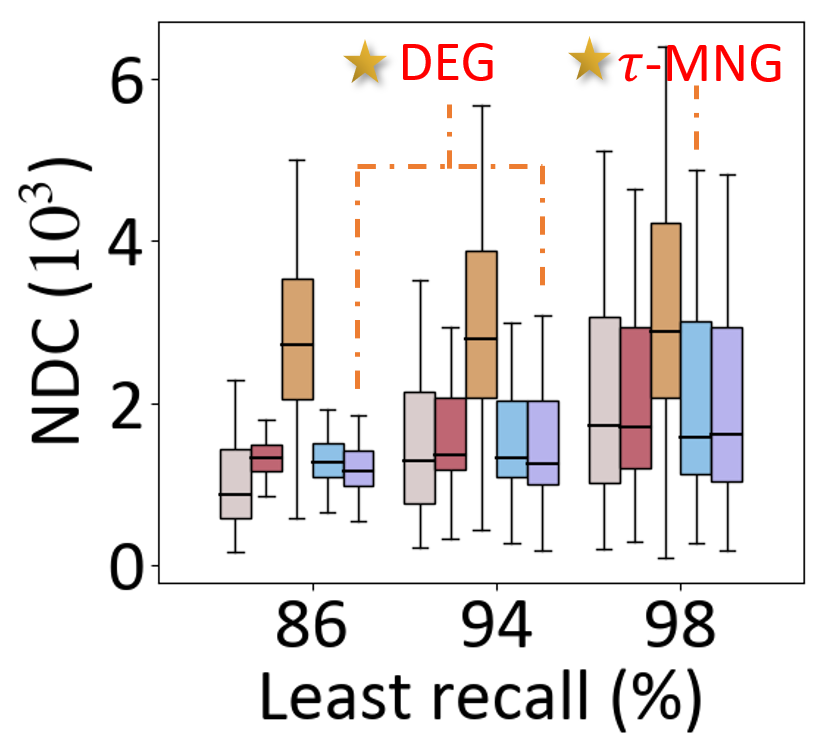}
}
\subfigure[Glove]{
\includegraphics[width=0.31\linewidth]{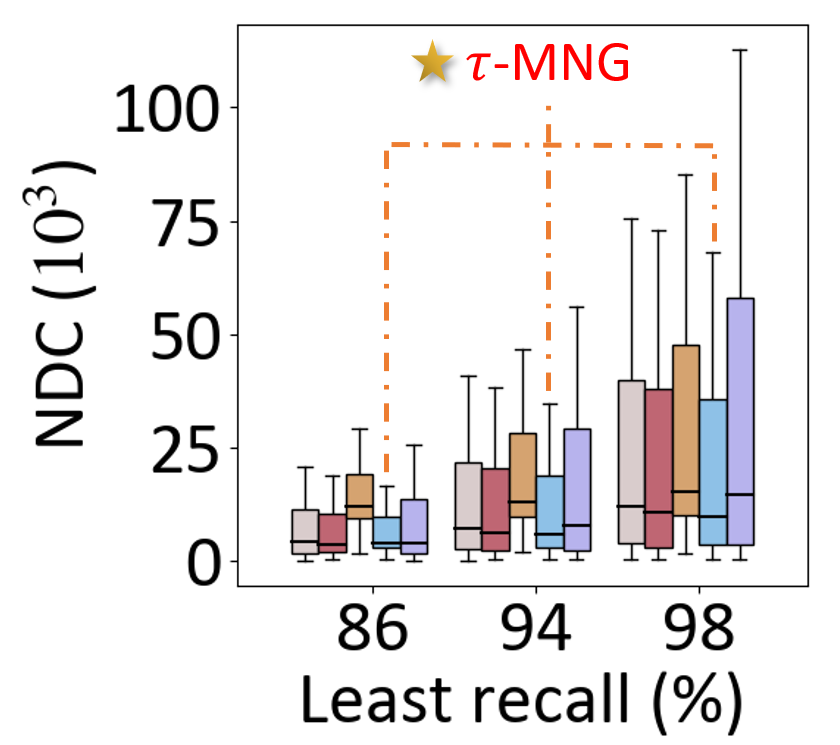}
}
\subfigure[Rand]{
\includegraphics[width=0.31\linewidth]{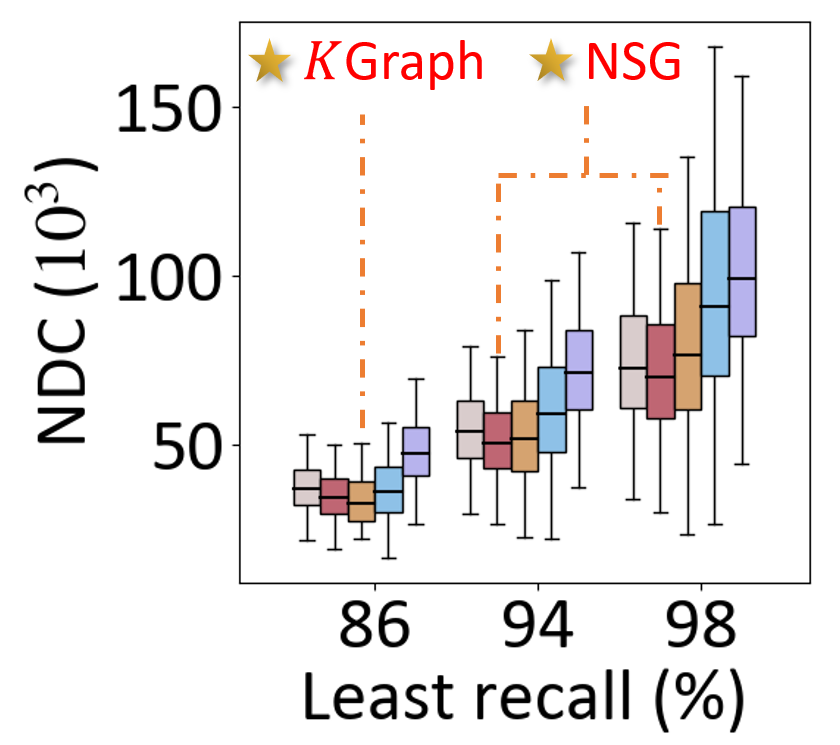}
}
\caption{Evaluate indexes on original query workloads.}
\label{fig:benchmark-old}
\end{minipage}
\begin{minipage}{0.475\linewidth}
\subfigure[Deep]{
\includegraphics[width=0.31\linewidth]{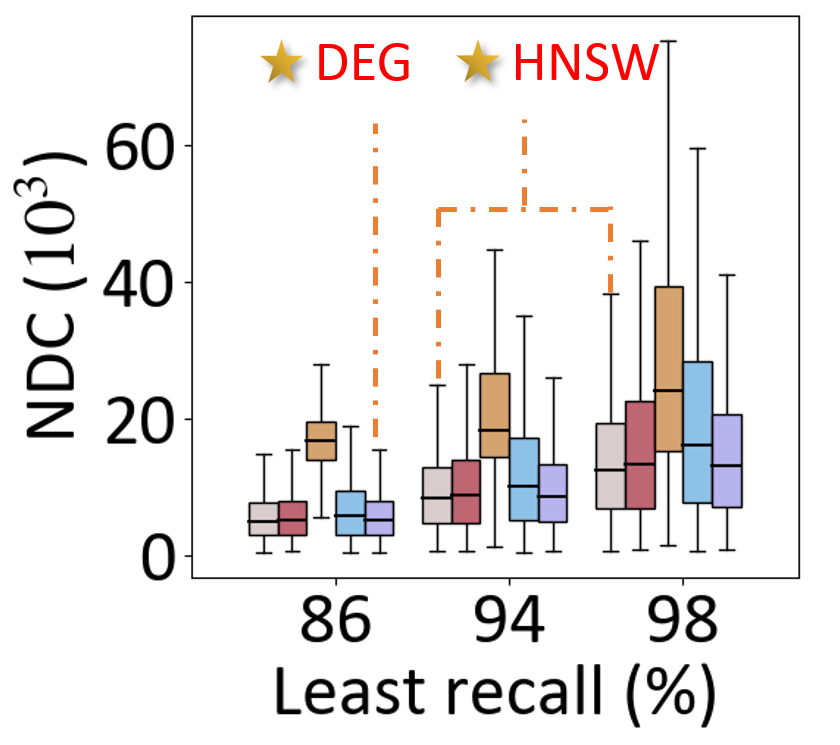}
}
\subfigure[Glove]{
\includegraphics[width=0.31\linewidth]{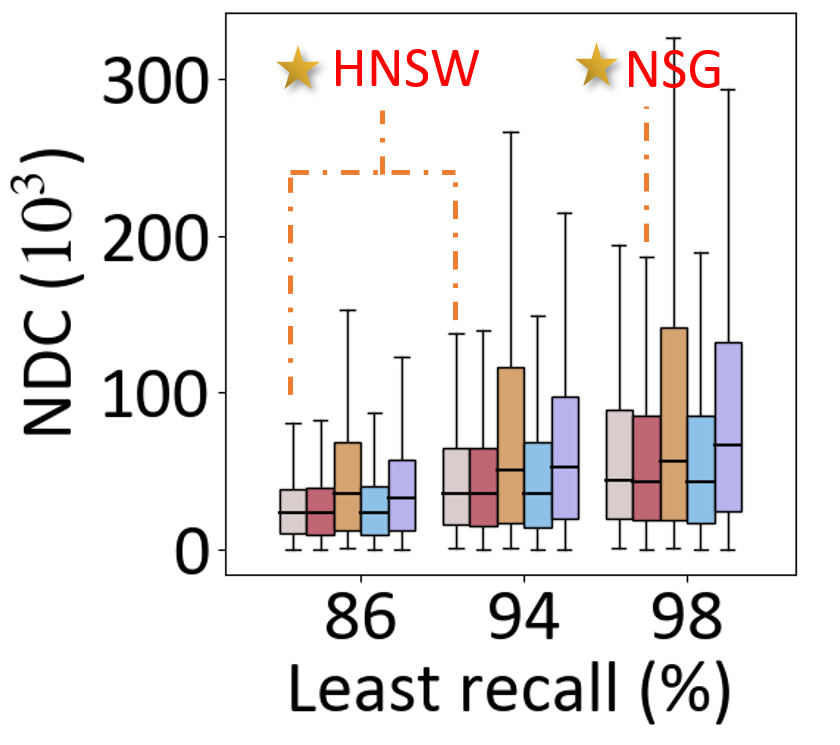}
}
\subfigure[Rand]{
\includegraphics[width=0.31\linewidth]{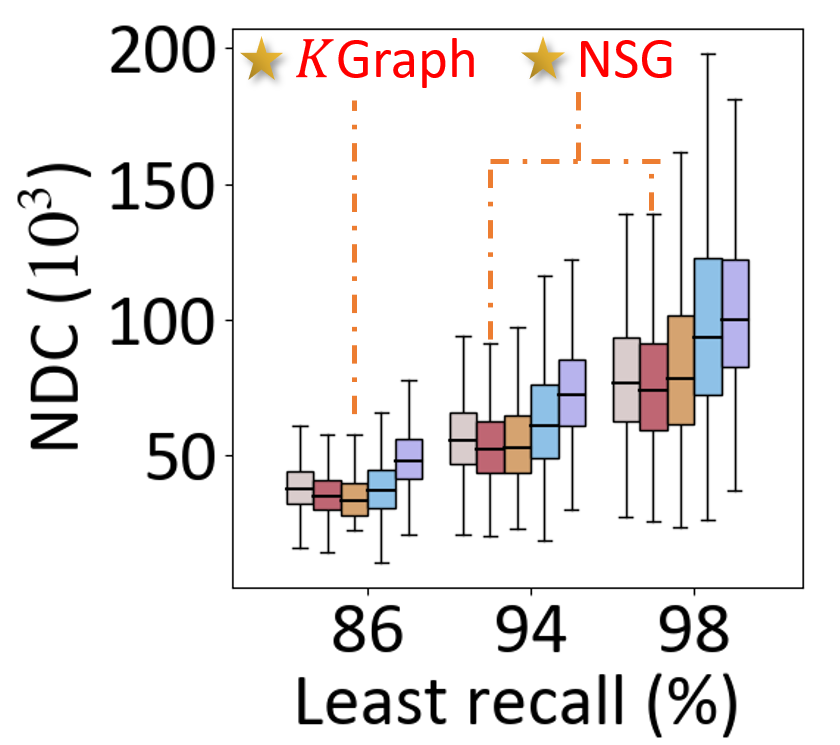}
}
\caption{Evaluate indexes on unbiased query workloads.}
\label{fig:benchmark}
\end{minipage}
\end{figure*}
\subsection{Effectiveness of $\boldsymbol{Steiner}$-hardness}

We then study the effectiveness of our index-independent $Steiner$-hardness.
That is, whether a query with larger $Steiner$-hardness requires more effort to be answered for graph indexes.
As studied in~\cite{bulletin-wang,stronger}, the insertion order when building graph indexes significantly impacts the effort to answer a given query.
Therefore, to mitigate randomness, we shuffle the dataset and build multiple indexes with different insertion orders on the same dataset, and take the average query effort on these indexes as the actual effort.

Figure~\ref{fig:hardness} shows the correlation coefficients between the query effort and different hardness measures on different datasets and indexes\footnote{DEG fails to be built on Tiny10M due to logic errors.}.
Our $Steiner$-hardness shows the strongest correlation among all the hardness measures, with on average \textbf{0.75} correlation coefficient, whereas $\epsilon$-hardness, LID, QE and RC are \textbf{0.50}, \textbf{0.42}, \textbf{0.26} and \textbf{0.31}, respectively, far smaller than $Steiner$-hardness.
These results demonstrate that $Steiner$-hardness can effectively represent the intrinsic hardness of the query on graph indexes.

It is observed that the correlation on real datasets is stronger than synthetic datasets.
It might because that real datasets are often distributed in a clustered manner~\cite{hub,hub2} and the $k$NN of the query are often highly interconnected in the graph~\cite{bulletin-wang}, which indicates that they will be visited together when querying.
In this way, the number of close points correlates more to the actual query effort than in the synthetic datasets.

\subsection{Comparison of Old and New Workloads}
\label{sec:expr-distrib}
We first check the hardness distribution of the queries in our newly generated workloads which are produced by adopting the approach described in Section~\ref{sec:workload-gen}
The new workload contains 1,000 queries, constructed by randomly selecting 50 queries from 20 hardness segments, i.e., $h$=20.
The result is shown in Figure~\ref{fig:dist}, where we compare our new unbiased workload against the original workload on three representative datasets\footnote{The box plot is shown in the most popular way with outliers~\cite{box-plot}.}.
Assume simple queries are the ones with hardness in the 20\% lowest range of the whole hardness spectrum.
Then the percentage of simple queries on the original workload is \textbf{78\%}, \textbf{80\%}, and \textbf{82\%} on the three datasets respectively, while \textbf{20\%} on the new unbiased workloads. 
This verifies our workloads do not show bias for simple or hard queries.

In Figure~\ref{fig:recall-diff}, we compare the query accuracy under old and new workloads using HNSW.
To fairly display the difference, we set NDC of each query as a fixed number and calculate the recall.
In the original workloads, the simple queries which cover the most of the workload, can quickly be answered with a high recall.
While the recall of hard queries is gradually improved as NDC increases.
On the contrary, in our unbiased workload, when NDC is small, the recall values of the queries are uniformly distributed.

\needspace{8mm}
\subsection{Index Evaluation on New Workloads}
To comprehensively show the query performance, we design a new metric that shows the distribution of NDC when all the queries in the workload reach a certain recall in Figure~\ref{fig:benchmark-old} and ~\ref{fig:benchmark}, for the old and the new unbiased workloads, respectively.
The best method w.r.t. the median value under each setting is marked with a star.
We summarize the major insights as follows.

\noindent(1) 
\emph{The performance significantly downgrades when using new unbiased workloads compared to the old one on real datasets.}
The NDC range (in y-axis) increases \textbf{>10} and \textbf{>3} times on Deep and Glove datasets respectively, from Figure~\ref{fig:benchmark-old} to ~\ref{fig:benchmark}.

\noindent(2) 
\emph{As the recall target increases, the performance variance of graph indexes goes larger.}
For example, on Deep dataset and HNSW index, after removing outliers, when recall=86\%, the range of NDC spans from 491 to 14,765, and 718 to 38,288 when recall=86\%.
This indicates that high recall targets can lead to a significant effort for the graph index, as has been observed before~\cite{note,stronger}.

\noindent(3) 
\emph{Latest indexes show improvement on old workloads, but can be beaten by HNSW or NSG on unbiased workloads.}
In Figure~\ref{fig:benchmark-old}, $\tau$-MNG or DEG wins the first or second place on most settings on real datasets even with our new metric.
While in Figure~\ref{fig:benchmark}, they become inferior to HNSW and NSG w.r.t. the median value or the variance.
This indicates that latest indexes might ``overfit'' to the old biased workloads, while the classical HNSW and NSG remain in the pool of graph index choices.

\noindent(4) \emph{Advanced indexes do not always outperform $K$Graph.}
On some settings of Rand and Tiny10M datasets, $K$Graph is a strong competitor.
Although the number of edges in $K$Graph is 2.5x more than the other graphs to reach the best performance, its better navigability leads to a superior overall query performance. 

\begin{figure}[tb]
\includegraphics[width=\linewidth]{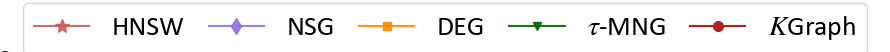}

\subfigure[Deep (simple)]{
\label{fig:deep-easy} 
\includegraphics[width=0.47\linewidth]{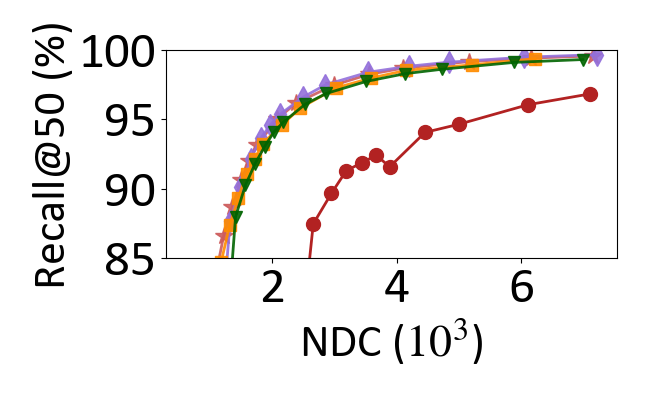}
}
\subfigure[Deep (hard)]{
\label{fig:deep-hard} 
\includegraphics[width=0.47\linewidth]{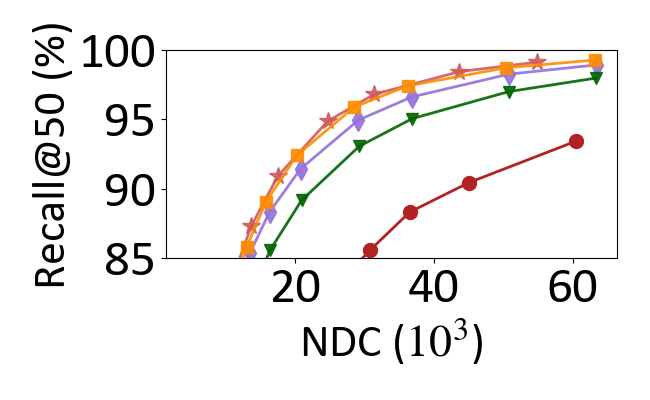}
}

\subfigure[GIST (simple)]{
\label{fig:gist-easy} 
\includegraphics[width=0.47\linewidth]{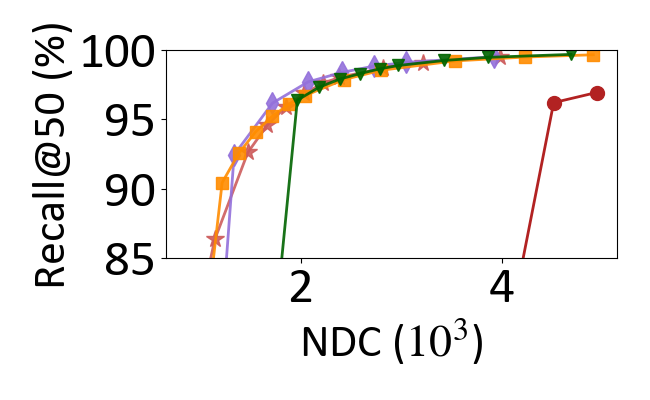}
}
\subfigure[GIST (hard)]{
\label{fig:gist-hard} 
\includegraphics[width=0.47\linewidth]{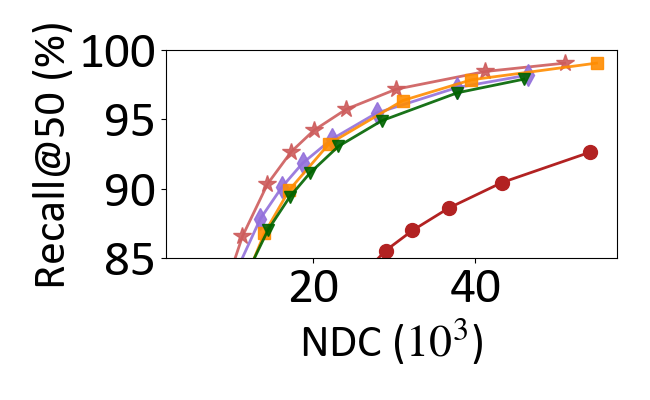}
}

\subfigure[Glove (simple)]{
\label{fig:glove-easy} 
\includegraphics[width=0.47\linewidth]{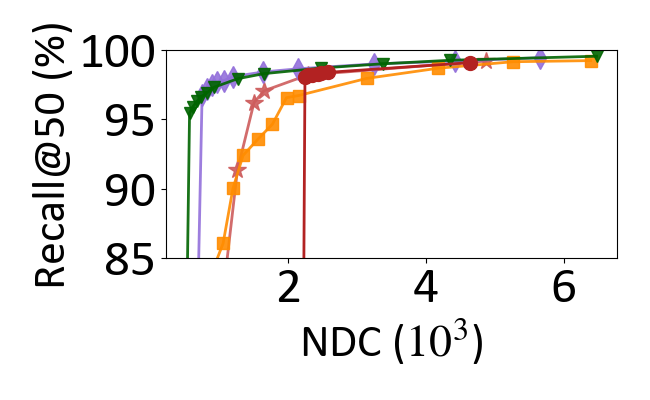}
}
\subfigure[Glove (hard)]{
\label{fig:glove-hard} 
\includegraphics[width=0.47\linewidth]{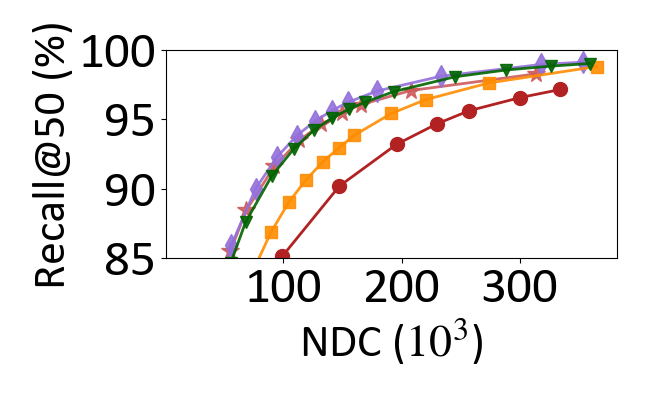}
}
\caption{Benchmark results on 1,000 simple and hard queries from our unbiased workloads, respectively.}
\label{fig:easy-hard-benchmark}
\end{figure}

\subject{Performance comparison between simple and hard queries.}
We select 1,000 simplest and hardest queries from the new workload and benchmark the indexes.
As shown in Figure~\ref{fig:easy-hard-benchmark}, even though all the indexes can answer hard queries, the time cost is higher than simple queries from \textbf{10} to \textbf{50} times (see the differences of the x-axis).
For the ranking result of simple queries, all indexes perform well on three datasets, except for $K$Graph on Deep and GIST.
While for hard queries, $\tau$-MNG and DEG show unstable performance on some cases. 
HNSW is the most robust followed by NSG. 

\subject{Discussion.}
According to our evaluation, the major problem of current graph indexes is the stability on hard queries.
The huge performance variance hinders their effects in actual applications.
$K$Graph surpringly works well on hard datasets, demonstrating that current edge pruning rules might not work well all the time.
Besides that, since the same hard query could be answered with variable efficiency on different index instances, combining multiple index instances is also a promising idea to improve the reachability of the graph.

\subsection{Ablation Study of ME}
\label{sec:expr-ablation}
\begin{figure}[tb]
\subfigure[Eventual performance]{
\label{fig:ablation1} 
\includegraphics[width=0.47\linewidth]{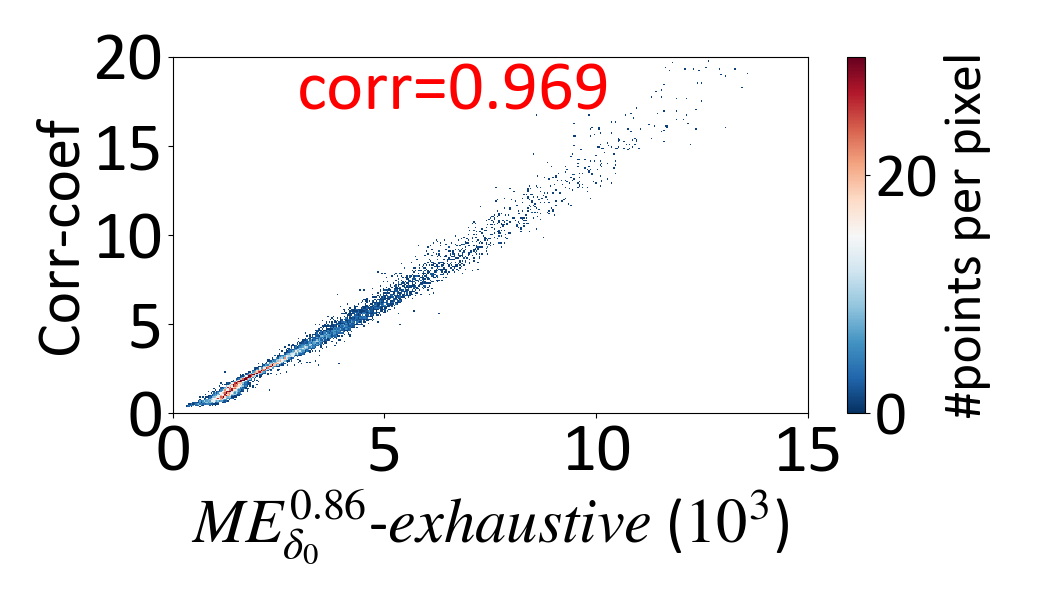}
}
\subfigure[Arbitrary entry point]{
\label{fig:ablation2} 
\includegraphics[width=0.47\linewidth]{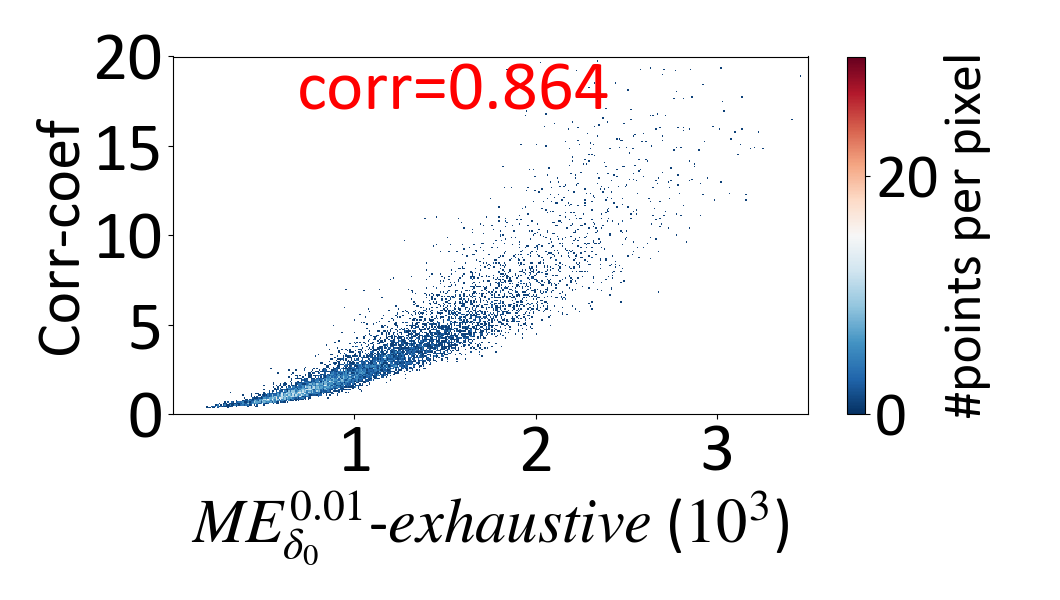}
}

\subfigure[Unlimited range of candidates]{
\label{fig:ablation3} 
\includegraphics[width=0.47\linewidth]{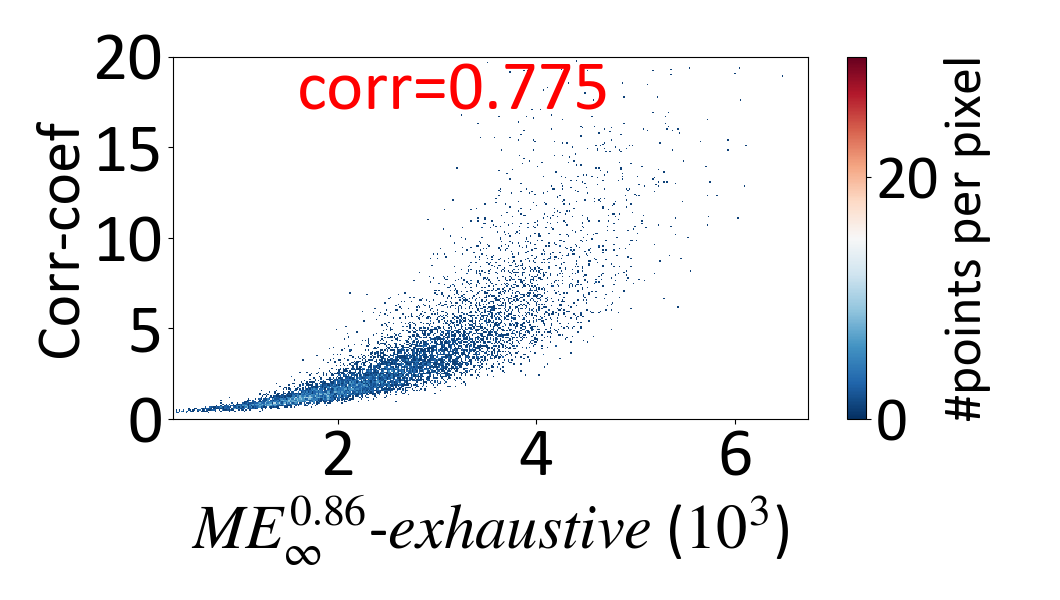}
}
\subfigure[Remove decision cost]{
\label{fig:ablation4} 
\includegraphics[width=0.47\linewidth]{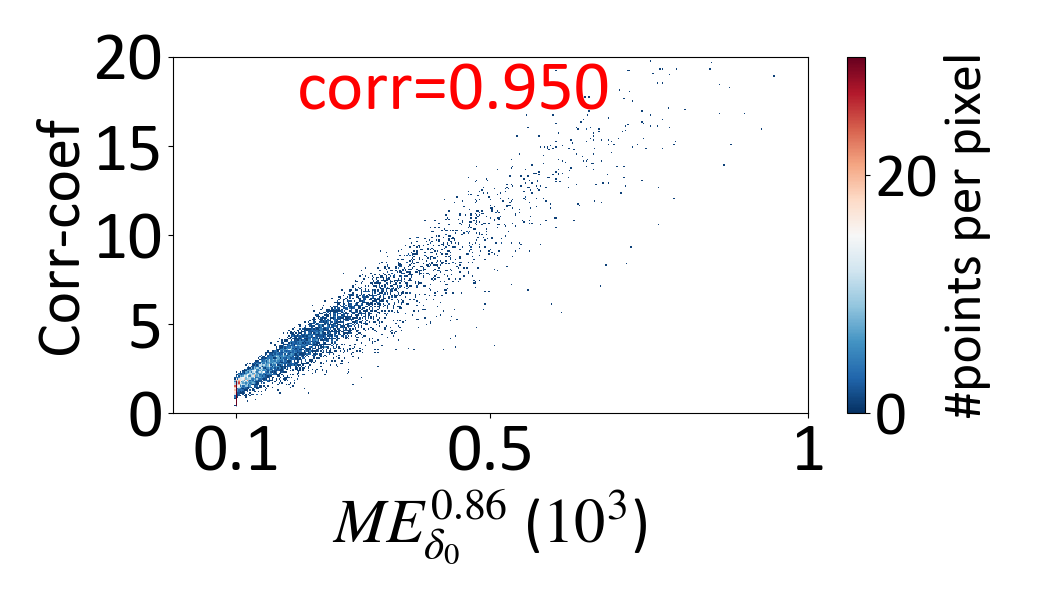}
}
\caption{Ablation studies on the definitions of ME.}
\label{fig:ablation}
\end{figure}

In Figure~\ref{fig:ablation}, we present ablation study results to understand the effect of the three constraints of $ME$ in Section~\ref{sec:cons-me}.
Figure~\ref{fig:ablation} (b)-(d) show the deterioration of the correlation between the estimated effort and the actual query effort.
We first remove the constraint on entry points (Constraint \ding{172}) by setting $p=1/k$, that is, using the optimal entry point in $N_k$ to get the ME.
As shown in Figure~\ref{fig:ablation2}, the correlation becomes 
weak 
for hard queries.
Figure~\ref{fig:ablation3} shows a similar result where the limitation on the accessible candidates (Constraint \ding{173}) is removed.
When removing the decision cost (i.e., Definition~\ref{def:me}), as shown in Figure~\ref{fig:ablation4}, simple queries cannot be distinguished.
In brevity, the three proposed constraints are all necessary to describe the query effort. 




\begin{table}[tb]
\footnotesize
\caption{One-off pre-processing time of Steiner-hardness}
\label{tab:pre-process}
\begin{tabular}{ccccccc}
\toprule
     & Deep    & GIST & Rand &  Glove & Tiny10M & Deep1B   \\ \midrule
Time (min) & 0.9 & 3.4  &  9.6    &   66.8  &   34.5    & 16.9 hrs\\ \bottomrule
\end{tabular}
\end{table}

\begin{figure}[tb]
\subfigure[Million-sized datasets]{
\label{fig:hardness-time}
\includegraphics[width=.475\linewidth]{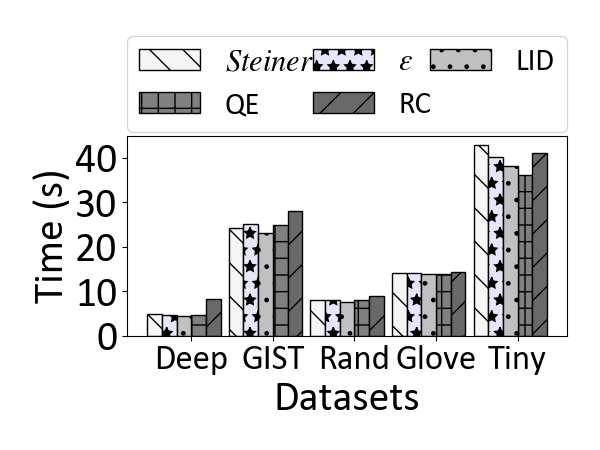}
}
\subfigure[Deep1B]{
\label{fig:hardness-scalability}
\includegraphics[width=.475\linewidth]{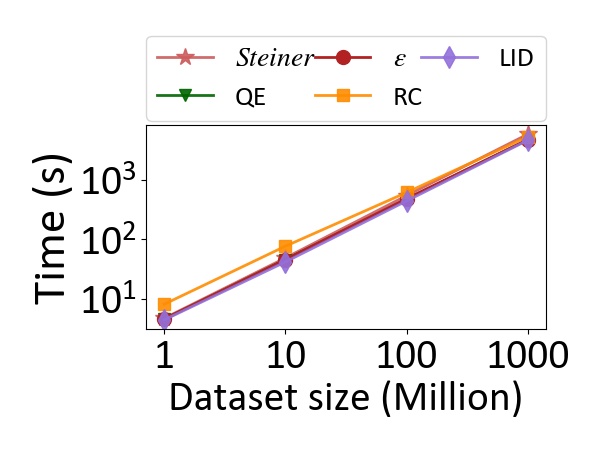}
}
\caption{The time of calculating the hardness 1,000 queries.}
\end{figure}

\subsection{Efficiency and Scalability}
We finally test the efficiency and the scalability of our approach.

\subject{One-off pre-processing.}
To calculate $Steiner$-hardness, for each dataset, we need to build the approximate MRNG and its reversed graph as pre-processing.
This one-off pre-processing time is reported in Table~\ref{tab:pre-process}.
This time is dominated by building a $K$Graph (>90\%) while building MRNG from the $K$Graph is very fast.

\subject{Hardness calculation.}
We report the overall time of calculating the $Steiner$-hardness of 1,000 queries in Figure~\ref{fig:hardness-time}, and compare it against the baselines on all the datasets.
It is observed that on five million-sized datasets, the time to compute different hardness measures is very close except for RC.
Figure~\ref{fig:hardness-scalability} shows all the hardness measures can be computed with a linear scalability.

\subject{Workload generation.}
Generating the new unbiased workload includes three steps: (1) GMM training and inference, (2) hardness calculation, and (3) query selection.
The first step can be done on a million-sized sample or smaller, which costs several to tens of minutes.
The time cost of the second step is shown in Figure~\ref{fig:hardness-time} while the third step only costs a constant time.

\section{Conclusions and Future Work}
\label{sec:con}
In this paper, we propose a practical query effort analyses framework for graph-base ANN indexes.
This framework effectively describes the minimum effort under different recall targets and $k$ on a given graph.
We further design a novel connection-based query hardness measure, $Steiner$-hardness, based on the framework for graph indexes with efficient algorithms.
Moreover, we build unbiased workloads encompassing the entire spectrum of $Steiner$-hardness to fairly stress-test current graph-based indexes. 

In our future work, we plan to conduct a more comprehensive evaluation with unbiased query workloads to benchmark both graph-based and partition-based indexes, and recommend indexes based on different user requirements.

\clearpage

\bibliographystyle{ACM-Reference-Format}
\bibliography{main}

\end{document}